\documentclass[12pt]{article}
\usepackage{comment}
\usepackage{amsmath}
\usepackage{amssymb}
\usepackage{graphicx}
\usepackage{here}
\usepackage{subcaption}
\usepackage{url}
\usepackage[sort&compress, numbers, merge]{natbib}
\usepackage{braket}
\usepackage{physics}
\usepackage[compat=1.1.0]{tikz-feynhand}
\usepackage{slashed}
\usepackage{mathtools}

\usepackage{placeins}

\numberwithin{equation}{section}

\usepackage[colorlinks=true, linkcolor=blue, citecolor=blue,
urlcolor=black]{hyperref}

\begin{document}
\def\meV{\mathrm{meV}}
\def\ps{\mathbf{p}}
\def\PS{\mathbf{P}}
\baselineskip 0.6cm
\def\simgt{\mathrel{\lower2.5pt\vbox{\lineskip=0pt\baselineskip=0pt
           \hbox{$>$}\hbox{$\sim$}}}}
\def\simlt{\mathrel{\lower2.5pt\vbox{\lineskip=0pt\baselineskip=0pt
           \hbox{$<$}\hbox{$\sim$}}}}
\def\simprop{\mathrel{\lower3.0pt\vbox{\lineskip=1.0pt\baselineskip=0pt
             \hbox{$\propto$}\hbox{$\sim$}}}}
\def\Im{\mathrm{Im}}
\def\calN{\mathcal{N}}
\def\Re{\mathrm{Re}}
\def\SU{\mathrm{SU}}
\def\tr{\mathrm{tr}}
\def\UFN{\mathrm{U(1)_{FN}}}
\def\ubar{\overline{u}}
\def\dbar{\overline{d}}
\def\ebar{\overline{e}}
\def\Nbar{\overline{N}}
\def\calZ{\mathcal{Z}}
\def\calL{\mathcal{L}}
\def\calO{\mathcal{O}}
\def\quark{\mathrm{quark}}
\def\dimfive{\mathrm{dim\mathchar`-5}}
\def\seesaw{\mathrm{seesaw}}
\def\P{\mathrm{P}}
\def\GUT{\mathrm{GUT}}
\def\CKM{\mathrm{CKM}}
\def\PMNS{\mathrm{PMNS}}

\begin{titlepage}

\begin{flushright}
IPMU24-0047
\end{flushright}

\vskip 1.1cm

\begin{center}

{\large \bf 
Comprehensive Bayesian Exploration
of Froggatt-Nielsen Mechanism
}

\vskip 1.2cm

\vskip 1.2cm
Masahiro Ibe$^{a,b}$, 
Satoshi Shirai$^{b}$ and
Keiichi Watanabe$^{a}$
\vskip 0.5cm

{\it

$^a$ {ICRR, The University of Tokyo, Kashiwa, Chiba 277-8582, Japan}

$^b$ {Kavli Institute for the Physics and Mathematics of the Universe
(WPI), \\The University of Tokyo Institutes for Advanced Study, \\ The
University of Tokyo, Kashiwa 277-8583, Japan}

}

\vskip 1.0cm

\abstract{
The Froggatt-Nielsen (FN) mechanism successfully explains the hierarchical structure of fermion Yukawa couplings by introducing a U(1) flavor symmetry with distinct charge assignments for different fermion generations.
While some FN charge assignments have been proposed, 
their evaluation has largely relied on heuristic approaches.
This paper systematically investigates viable FN charge assignments within the Standard Model, 
including both the quark and lepton sectors, 
using Bayesian statistical analysis. 
The study explores scenarios involving both the seesaw mechanism and dimension-five operators for neutrino mass generation. 
A comprehensive parameter scan over FN charges reveals a wide range of charge assignments consistent with observed fermion masses and mixing angles. 
Interestingly, 
negative FN charges and significant generational differences in charges are found to be viable, 
contrary to conventional assumptions. 
The analysis also compares the seesaw mechanism and dimension-five operator scenarios,
finding no strong preference between them for optimal charge assignments. 
Furthermore,
predictions for the lightest neutrino mass and effective Majorana mass relevant for neutrinoless double-beta decay are presented, 
highlighting regions of parameter space accessible to upcoming experiments.
Finally,  
implications for nucleon decay are studied, demonstrating that different FN charge assignments predict significantly different nucleon decay lifetimes and branching ratios, 
providing a potential experimental probe for FN models. 
}

\end{center}
\end{titlepage}

\tableofcontents

\section{Introduction} 

Despite the great success of the Standard Model (SM), 
several puzzles remain.
One of them is the flavor puzzle. 
If the SM is a natural theory which 
consists of dimensionless parameters of $\order{1}$, 
all the SM fermions should have similar masses.
All the mixing angles between the fermions should also be of $\order{1}$.
In reality, 
however, 
there are hierarchies between the angles and the fermion masses.
For example, 
the quark mixing angles are ordered as
\begin{align}
    \mathrm{sin}^{4}\, \theta_{12}^{\CKM} \sim \mathrm{sin}^{2}\, \theta_{23}^{\CKM}
    \sim \mathrm{sin}\, \theta_{13}^{\CKM} \, ,
\end{align}
with $\sin\theta_{12}^{\CKM}\simeq 0.22$.
Besides, 
there are intriguing relations between the quark mass hierarchy and the mixing angles,
\begin{align}
    & \mathrm{sin}\, \theta_{12}^{\CKM} \sim \sqrt{m_d/m_s}
    \sim \sqrt{m_s/m_b}  \, , \\ 
    &\mathrm{sin}\, \theta_{23}^{\CKM} \sim \sqrt{m_u/m_c} 
    \sim \sqrt{m_c/m_t} \, , \\
    &\mathrm{sin}\, \theta_{13}^{\CKM} \sim \sqrt{m_u/m_t} 
    \sim \sqrt{m_d/m_b} \, .
\end{align}
These relationships seem to suggest the existence of an underlying flavor structure.
The charged lepton masses also have similar hierarchy to the down type quark masses.
In the neutrino sector, on the other hand, 
the mixing angles are much less hierarchical, which makes the flavor puzzle more interesting.
To explain the patterns and characteristics of these flavor structures, various models and ideas have been proposed (see e.g., Refs.\,\cite{Babu:2009fd,Xing:2020ijf} for reviews and references therein).
Among them, the Froggatt-Nielsen (FN) mechanism\,\cite{Froggatt:1978nt} based on a U(1) symmetry, 
$\UFN$, 
is the simplest model.

The FN mechanism can construct hierarchical Yukawa coupling structures by utilizing a U(1) symmetry and its subsequent breaking, 
even when the fundamental dimensionless couplings are of $\order{1}$. 
At its core, 
the FN mechanism depends on the U(1) charge assignments on matter fields, 
making the identification of appropriate charge assignments the most critical task in model building. 
However, 
due to the uncertainties of $\order{1}$ couplings, 
it is challenging to resolve the degeneracy of possible charge assignments that can reproduce the flavor structure. 
Numerous studies have attempted to identify suitable charge assignments (see e.g., Refs.\,\cite{Leurer:1992wg,Leurer:1993gy}; for a review, 
see Refs.\,\cite{Altmannshofer:2022aml}). 
Most of these studies, 
however, 
lack an objective framework for evaluating the ``goodness'' of FN charge assignments. 
An objective assessment of the ``goodness'' of charge assignments for the lepton sector was explored in Ref.\,\cite{Bergstrom:2014owa} using Bayesian analysis. 
Nevertheless, 
there has been no systematic study of generic charge assignments that includes both the quark and lepton sectors simultaneously.

In light of these challenges, the primary goal of this paper is as follows.
By extending the analysis in Ref.\,\cite{Bergstrom:2014owa},
we perform Bayesian analysis
and conduct a comprehensive exploration of the FN charge parameter space with absolute values less than or equal to $10$.
In our analysis, 
we first discuss the FN charge assignments for the quark and the lepton sector
separately.
For the lepton sector, 
we consider both the seesaw mechanism and the dimension-five operator as origins of neutrino masses and discuss the preferred origin.
Additionally,
we perform simultaneous Bayesian analysis of quark and lepton charge assignments, 
which provides implications for the FN charge assignments within the Grand Unified Theory (GUT).

The FN mechanism may affect phenomena in flavor-changing processes which are not described in the SM.
Among those phenomena,
we discuss the implication of nucleon decay caused by dimension-six effective operators under the FN mechanism.
We find that there are numerous FN charge assignments that explain the flavor structure equally well in this paper, 
but the behavior of nucleon decay varies significantly depending on the FN charges.
Thus, 
the observation of nucleon decay could serve as a powerful probe to investigate the origin of the flavor structure.

The organization of this paper is as follows.
In Sec.\,\ref{sec:FNmechanism}, 
we provide an overview of the FN mechanism. 
In Sec.\,\ref{sec:inference},
we outline the Bayesian inference framework that offers a quantitative criterion for evaluating the efficacy of FN charge assignments. 
In Sec.\,\ref{sec:result_charge},
we apply this Bayesian inference to FN charge assignments and compare various models.
In Sec.\,\ref{sec:mbb}, 
we discuss the probability distribution function of the lightest neutrino mass, $m_1$ for a ``good" FN charge assignment.
We also present a prediction of the effective Majorana mass of $\nu_e$, $m_{ee}$ which is related to the neutrinoless double beta decay.
In Sec.\,\ref{sec:neuleon_decay},
we discuss the branching fractions of various nucleon decay mode under the FN mechanism. 
The final section is devoted to our conclusions.

\section{Froggatt-Nielsen Mechanism}
\label{sec:FNmechanism}

\subsection{Yukawa couplings in Standard Model}
\label{sec:yukawa}

The SM Yukawa interactions are 
\begin{align}
\label{eq:SM yukawa}   \mathcal{L}_{\mathrm{Y}} 
   = - y^{(u)}_{ij}\, Q_{i}\, \ubar_{j} H 
   - y^{(d)}_{ij}\, Q_{i}\, \dbar_{j} H^{\dagger}
   - y^{(e)}_{ij}\, L_{i}\, \ebar_{j} H^{\dagger}
   + h.c.
\end{align}
Here, $H$ denotes the Higgs doublet,
$Q =(u,d)$, $\bar{u}$, $\bar{d}$ 
are the doublet, the anti-up-type, and the anti-down-type quarks, while $L=(\nu,e)$, 
$\bar{e}$ are the doublet and the anti-electron-type leptons
with $i,j$ being flavor indices.%
\footnote{In this paper, we use Weyl fermion notation thoroughly with the notation used in Ref.\,\cite{Dreiner:2008tw}.
}
The convention for the 
SM charges is the same as 
those in Ref.\,\cite{Dreiner:2008tw}.
The Yukawa coupling constants, $y^{(u,d,e)}$, 
are each a $3\times 3$ matrix of complex numbers.

With vacuum expectation value (VEV)
of the Higgs doublet, $\langle H \rangle = (0,v_\mathrm{EW})^T$,%
\footnote{We assume $v_\mathrm{EW}>0$, which does not lose generality.
}
the quarks and the charged leptons obtain masses,
\begin{align}
   \mathcal{L}_{\mathrm{mass}} 
   = - m_{ij}^{(u)}\, u_{i}\, \ubar_{j} 
   - m_{ij}^{(d)}\, d_{i}\, \dbar_{j}
   - m_{ij}^{(e)}\, e_{i}\, \ebar_{j}
   + h.c.,
\end{align}
where 
\begin{align}
&m^{(u)}_{ij} 
= 
y^{(u)}_{ij} v_\mathrm{EW}\ , \,\quad 
m^{(d)}_{ij} 
= 
y^{(d)}_{ij} v_\mathrm{EW}\ , \,\quad 
m^{(e)}_{ij}
= 
y^{(e)}_{ij} v_\mathrm{EW}\ .
\end{align}
These mass matrices can be diagonalized by using two unitary matrices as follows,
\begin{align}
\label{eq:SVD of fermions}
&m^{(u)}_{ij} 
= 
U_{uL,ik} \,\, 
\hat{m}^{(u)}_{k} \, 
U^{\dagger}_{uR,kj}\ , 
\quad  \,\, 
m^{(d)}_{ij} 
= 
U_{dL,ik} \,\, 
\hat{m}^{(d)}_{k} \, 
U^{\dagger}_{dR,kj}\ ,
\,\quad
m^{(e)}_{ij} 
= 
U_{eL,ik} \,\, 
\hat{m}^{(e)}_{k} \, 
U^{\dagger}_{eR,kj}\ .
\end{align}
Here, $\hat{m}_{k}$'s represent the diagonalized mass matrix, with its diagonal components being positive valued. 
The diagonal components are arranged in ascending order.
Due to the noncommutativity between
mass diagonalization and the SU(2)$_L$ gauge interaction,
the CKM matrix arises in the $W$-boson interaction in the quark sector,
\begin{align}
V_{\CKM}
=
U_{uL}^{T} U_{dL}^{*}\ .
\end{align}

\subsection{Interactions in neutrino sector}
In this paper, we assume that neutrinos have Majorana masses. 
Regarding their origin, we examine two scenarios: the seesaw mechanism~\cite{Minkowski:1977sc,Yanagida:1979as,*Yanagida:1979gs,Gell-Mann:1979vob,Glashow:1979nm,Mohapatra:1979ia} and general dimension-five operators~\cite{Weinberg:1979sa}.
In the case of the seesaw mechanism,
neutrino masses are generated from
the Yukawa interactions to the right-handed neutrinos,
\begin{align}
\label{eq:seesaw_LY}
\mathcal{L}_{\nu} 
= - y^{(D)}_{ia}\, L_i\, \Nbar_{a} H
- \frac{1}{2}\, M_{R}\, y_{ab}^{(R)}\, \Nbar_{a}\, \Nbar_{b} + h.c.  
\end{align}
Here, $\Nbar_a$ denote right-handed neutrinos with 
$a,b=1,2,3$ being flavor index.
$M_R\,(>0)$ is the mass scale of the right-handed neutrinos.
The Yukawa coupling constants,
$y^{(D)}$ and $y^{(R)}$, are a
$3\times3$ complex matrix and a $3\times3$ complex symmetric matrix, respectively.
By integrating out the right-handed neutrinos,
the dimension-five operators
appear as,
\begin{gather}
\label{eq:seesaw_dim5}
\mathcal{L}_{\nu} 
= 
- y^{(S)}_{ij}\,
\frac{(L_i H) (L_j H)}{2M_R} + h.c., \quad
y^{(S)}_{ij} 
= 
-(y^{(D)} (y^{(R)})^{-1}\, 
(y^{(D)})^T)_{ij} \ .
\end{gather}
With the Higgs VEV, the neutrinos obtain Majorana masses, 
\begin{align}
\mathcal{L}_{\mathrm{mass}}
=    - \frac{1}{2} m_{ij}^{(\nu)} \nu_{i} \, \nu_{j} + h.c.\, , \quad
m^{(\nu)}_{ij} 
= y^{(S)}_{ij} \frac{v^{2}_\mathrm{EW}}{M_R} \ .
\end{align}
The neutrino masses form a complex symmetric matrix. 
This matrix can be diagonalized using the Takagi decomposition with a single unitary matrix,
\begin{align}
m^{(\nu)}_{ij} 
= U_{\nu,ik} \,\, \hat{m}^{(\nu)}_{k} \, 
U^{T}_{\nu,kj}\ .
\end{align}
As in the case of the CKM matrix, the mass diagonalizations 
of the neutrinos and the charged leptons induce the PMNS matrix, 
\begin{align}
V_{\mathrm{PMNS}} = U_{eL}^{T} U_{\nu}^{*}\ .
\end{align}

In the second scenario, 
the neutrino masses 
arise from the dimension-five operators, 
\begin{gather}
\label{eq:wop}
\mathcal{L}_{\nu} 
= 
- y^{(W)}_{ij}
\frac{(L_i H) (L_j H)}{2\Lambda_W} + h.c. \, , 
\quad 
\end{gather}
where a dimensionful parameter $\Lambda_W > 0 $ denotes the cutoff scale and $y^{(W)}$ is a $3\times3$ complex symmetric matrix.
Then, the Majorana neutrino matrix is given by,
\begin{align}
  m^{(\nu)}_{ij} 
= y^{(W)}_{ij} \frac{v^{2}_\mathrm{EW}}{\Lambda_W}\ , 
\end{align}
which can be also diagonalized by the Takagi decomposition.
As we will see later, the Bayesian  analysis of the FN charges in the seesaw mechanism and the dimension-five operator yields different preferences.

\subsection{U(1) flavor model}
\label{sec:summaryFNmechanism}

The Froggatt-Nielsen mechanism is among the most successful models for explaining the hierarchy of the charged fermion masses and the quark mixing angles\,\cite{Froggatt:1978nt}. 
In this mechanism, a new U(1) flavor symmetry, U$(1)_{\mathrm{FN}}$, and a new complex scalar field $\Phi$ are introduced. 
Under this U(1) symmetry, quarks and leptons of different generations have distinct U$(1)_\mathrm{FN}$ charges, and $\Phi$ has an FN charge of $-1$. 
The U$(1)_{\mathrm{FN}}$ is spontaneously broken by the VEV of $\Phi$, $\langle\Phi\rangle$. Note that we assume U$(1)_{\mathrm{FN}}$ is not an exact global symmetry, and hence, we do not impose anomaly-free conditions on the FN charges. 
We also assume that the pseudo-Goldstone boson associated with the spontaneous U$(1)_{\mathrm{FN}}$ breaking obtains a sizable mass so that its dynamics does not significantly affect low-energy phenomenology, while the flavor structure remains intact despite the explicit U$(1)_{\mathrm{FN}}$ breaking.

The typical Yukawa interaction is prohibited due to U(1)$_\mathrm{FN}$ symmetry, and the interactions between fermions and scalars are generally proportional to the powers of $\Phi^{(\dagger)}$ as follows:
\begin{align}
\label{eq:LY}
 \mathcal{L}_{\mathrm{Y}} 
 = \kappa_{ij} \left(\frac{\Phi^{(\dagger)}}{M_{*}}\right)^{|f_{\psi,i} + f_{\chi,j}|} \psi_{i} \chi_{j} H^{(\dagger)} + h.c. 
\end{align}
Here, $M_{*}$ denotes the some cutoff scale, while 
$\kappa$ is a $3\times 3$ complex matrix.
$\psi_i$ and $\chi_i$ collectively denote $ Q_i, L_i$ and $\bar{u}_i$, $\bar{d}_i$, $\bar{e}_i$ (see Eq.\,\eqref{eq:SM yukawa}).
The $f_{\psi,i}$ and $f_{\chi,i}$ are the FN charges of the SM fermions, and we assume 
the Higgs boson does not have the FN charge.
Furthermore, 
since we assume a local effective theory, 
the effective Yukawa interactions involve the powers of 
$\Phi$ for $f_{\psi,i}+f_{\chi,j}>0$
and $\Phi^\dagger$ for $f_{\psi,i}+f_{\chi,j}<0$.

Once $\Phi$ develops a VEV, 
the FN symmetric terms in Eq.\,\eqref{eq:LY} result in the SM Yukawa interactions 
where the Yukawa couplings are given by 
\begin{gather}
\label{eq:hadamard}
   y^{(u)}_{ij} = \kappa^{(u)}_{ij} \circ \epsilon^{q^{(u)}_{ij}}\ , 
\quad
   y^{(d)}_{ij} = \kappa^{(d)}_{ij} \circ \epsilon^{q^{(d)}_{ij}}\ ,
\quad
   y^{(e)}_{ij} = \kappa^{(e)}_{ij} \circ \epsilon^{q^{(e)}_{ij}}\ .
\end{gather}
Here, $\circ$ represents the Hadamard product%
\footnote{The Hadamard product is the product of matrices determined by taking component-wise products over matrices of the same size.
}
and we do not sum over $i,j$ in Eq.\,\eqref{eq:hadamard}.
Also, FN breaking parameter, $\epsilon$, and $q_{ij}$'s are defined as follows:
\begin{gather}
\label{eq:delta}
   \epsilon \equiv \frac{\langle\Phi\rangle}{M_{*}} = \frac{\langle\Phi^{\dagger}\rangle}{M_{*}}\ , \\
\label{eq:chargeq}
   q^{(u)}_{ij} = |f_{Q,i} + f_{\ubar,j}|\ , 
\quad
   q^{(d)}_{ij} = |f_{Q,i} + f_{\dbar,j}|\ ,
    \quad
   q^{(e)}_{ij} = |f_{L,i} + f_{\ebar,j}|\ .
\end{gather}
Here, we have taken $\epsilon>0$ without loss of generality.
Since we assume $\epsilon < 1$, the above couplings are more suppressed for larger values of $q_{ij}$,
which generates non-trivial hierarchies of the Yukawa couplings even for $\kappa$'s of $\order{1}$.

The FN mechanism also affects the neutrino masses and mixing structure.
In the case of the seesaw mechanism,
the Yukawa coupling constants of the neutrino sector are given by
\begin{gather}
\label{eq:hadamard_nu}    
   y^{(D)}_{ia} = \kappa^{(D)}_{ia} \circ \epsilon^{q_{ia}^{(D)}}\ ,
   \quad
   y^{(R)}_{ab} = \kappa^{(R)}_{ab} \circ \epsilon^{q_{ab}^{(R)}}\ , \\
   q^{(D)}_{ia} = |f_{L,i}+f_{\Nbar,a}|\ ,
   \quad 
   q^{(R)}_{ab} = |f_{\Nbar,a}+f_{\Nbar,b}|\ .
\end{gather}
In the case of the dimension-five operators,
FN charge dependence of those is described by
\begin{gather}
\label{eq:kappaW}
   y^{(W)}_{ij} 
   = \kappa^{(W)}_{ij} \circ \epsilon^{q^{(W)}_{ij}}, \quad
    q^{(W)}_{ij} 
    = |f_{L,i} + f_{L,j}|\ .
\end{gather}
Here, $\kappa^{(D)}$ is 
a complex $3\times 3$ matrix
while $\kappa^{(R,W)}$
are complex $3\times 3$ symmetric matrices, which appear as 
the coefficients of the 
interactions involving the FN-breaking field as in \eqref{eq:LY}.
Once $\Phi$ develops a VEV, those 
interactions result in the terms
in Eq.\,\eqref{eq:seesaw_LY}
and Eq.\,\eqref{eq:wop}.
Note that we do not sum over $i,a,b$ in Eqs.\,\eqref{eq:hadamard_nu} 
and \eqref{eq:kappaW}.

\subsection{Heuristic approach to FN charge assignment}
\label{sec:heuristic_example}

In the conventional approach,
the FN breaking parameter $\epsilon$ in Eq.\,\eqref{eq:delta} is 
set to be around the Cabibbo angle, i.e., $\epsilon \sim 0.2$. 
The FN charges are chosen to approximately match the observed patterns of the quark and charged lepton masses as well as the CKM and PMNS angles. 
For example, 
they are given as
\begin{align}
\label{eq:charge_example}
f_Q=(3,2,0) \ , \quad 
f_{\ubar} = (5,2,0)\ , \quad
f_{\dbar} = (4,4,3)\ , \quad 
f_L = (4,3,3)\ , \quad
f_{\ebar} =(4,1,0)\ ,
\end{align}
where the numbers in parentheses represent the FN charge of each generation.
With these assumptions, the SM and the neutrino parameters are predicted to be, 
\begin{gather}
y_u : y_c : y_t
= 
\epsilon^8 : \epsilon^4 : \epsilon^0, \quad 
y_d : y_s : y_b
=
\epsilon^7 : \epsilon^6 : \epsilon^3, \quad
y_e : y_{\mu} : y_{\tau} 
=
\epsilon^8 : \epsilon^4 : \epsilon^3, \\
|V_{\CKM}|
\equiv 
\left( \begin{array}{ccc}
|V_{ud}| & |V_{us}| & |V_{ub}| \\
|V_{cd}| & |V_{cs}| & |V_{cb}| \\
|V_{td}| & |V_{ts}| & |V_{tb}|\end{array} \right) 
= 
\left( \begin{array}{ccc}
\epsilon^{0} & \epsilon^{1} & \epsilon^{3} \\
\epsilon^{1} & \epsilon^{0} & \epsilon^{2} \\
\epsilon^{3} & \epsilon^{2} & \epsilon^{0}
\end{array} \right), \\ |V_{\mathrm{PMNS}}|
\equiv 
\left( \begin{array}{ccc}
|V_{e1}| & |V_{e2}| & |V_{e3}| \\
|V_{\mu 1}| & |V_{\mu 2}| & |V_{\mu 3}| \\
|V_{\tau 1}| & |V_{\tau 2}| & |V_{\tau 3}|
\end{array} \right) 
= 
\left( \begin{array}{ccc}
\epsilon^{0} & \epsilon^{1} & \epsilon^{1} \\
\epsilon^{1} & \epsilon^{0} & \epsilon^{0} \\
\epsilon^{1} & \epsilon^{0} & \epsilon^{0}
\end{array} \right) \ .
\end{gather}
up to $\order{1}$ coefficients stemming from $\kappa$'s.

\begin{figure}[t!]
	\centering
 	\subcaptionbox{}
{\includegraphics[width=0.49\textwidth]{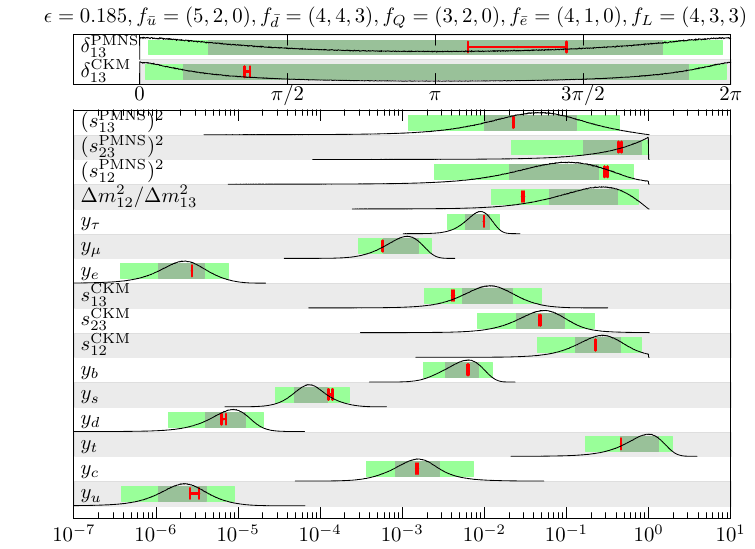}}
 	\subcaptionbox{}	{\includegraphics[width=0.49\textwidth]{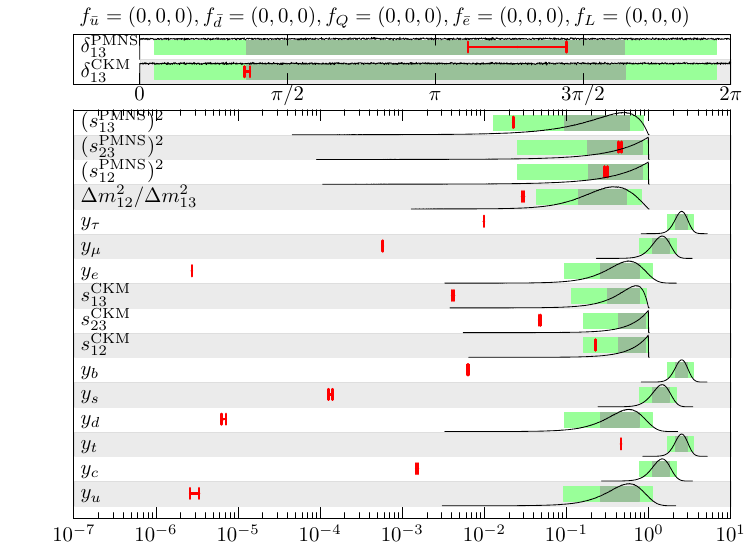}}
 \caption{(a) Predictions
 (i.e., the prior distributions)
 from $\order{1}$ distributions of $\kappa$'s with the FN charge assignment 
 in Eq.\,\eqref{eq:charge_example}. 
(b) Predictions from $\order{1}$ distributions of $\kappa$'s without the FN mechanism.
The red bars show the SM 
parameters in the 
$\mathrm{\overline{MS}}$ scheme given 
in the App.\,\ref{sec:msbar}.
The dark green and light green bands correspond to the $1\sigma$ and $2\sigma$ percentiles, respectively.
}
 \label{fig:conv}
\end{figure}

To see how well these charge assignments reproduce the observed flavor structure, 
we present the distributions of the predicted physical parameters assuming the $\order{1}$ coefficients $\kappa$'s 
follow the normal distribution in Fig.\,\ref{fig:conv}.
Specifically, we take
\begin{align}
\label{eq:distribution_1}
\Re\, \kappa^{(u,d,e,D)}_{ab}
= \frac{\sigma}{\sqrt{2}}\, \calN(0,1)\, ,
\quad \Im\, \kappa^{(u,d,e,D)}_{ab}
= \frac{\sigma}{\sqrt{2}}\, \calN(0,1)\, .
\end{align}
Here, 
$\calN(0,1)$ denotes a random number drawn from a normal distribution with a mean of $0$ and a standard deviation of $1$.
The ensemble average of each $\kappa_{ab}$ is $\langle \kappa_{ab} \rangle = 0 $ and
we choose the normalization of the distribution so that the standard deviation $\sqrt{\langle |\kappa_{ab}|^2 \rangle} = \sigma$.
Similarly, 
the Majorana neutrino mass parameters $\kappa^{(R)}$ and $\kappa^{(W)}$ are represented by a complex symmetric matrix, and we assume the following distributions:
\begin{gather}
\Re\, \kappa^{(R,W)}_{aa} 
= 
\frac{\sigma}{\sqrt{2}}\, \calN(0,1)\, , \quad
\Im\, \kappa^{(R,W)}_{aa}
= 
\frac{\sigma}{\sqrt{2}}\, \calN(0,1)\, , \quad \nonumber \\
\Re\, \kappa^{(R,W)}_{ab} 
= \frac{\sigma}{2}\, \calN(0,1)\,  , \quad
\Im\, \kappa^{(R,W)}_{ab}
= \frac{\sigma}{2}\, \calN(0,1)\, ,\quad (a> b)\ .
\label{eq:distribution_2}
\end{gather}
Note that we assume the variance of the off-diagonal elements is
smaller than that of the diagonal elements by a factor of $\sqrt{2}$,
makes the distributions flavor symmetric due to the Majorana nature of the neutrinos.

In Fig.\,\ref{fig:conv}, the left panel shows 
the predicted distributions 
(i.e., the prior distributions)
of the 
parameters
for the FN charge assignment 
in  Eq.\,\eqref{eq:charge_example} with $\epsilon = 0.185$ and 
$\sigma = 1$.
Here we consider the case that neutrino masses are given by dimension-five operators in Eq.\,\eqref{eq:wop}.
In addition, 
we assume the normal neutrino mass ordering, i.e., ($m_1 < m_2 < m_3$).
The right panel shows the parameter distributions for the $\order{1}$ hypothesis without FN mechanism,
that is, all the FN charges $f_{Q,\ubar,\dbar,L,\ebar}$ are zero.
The dark green and light green bands correspond to the $1\sigma$ and $2\sigma$ percentiles, respectively.
The red bars show the values of the running parameters which are estimated in the $\overline{\mathrm{MS}}$ scheme at $\mu_{R}=10^{15}$ GeV in Tab.\,\ref{tab:parameters} (see App.\,\ref{sec:msbar}).
Black lines indicate the distributions of prediction for each parameter.
From Fig.\,\ref{fig:conv}, we find that the observed physical parameters are within typical ranges of 
the predictions of the FN mechanism with
the FN charge in Eq.\,\eqref{eq:charge_example} (left-panel).
This result should be compared with the prediction in the model without the FN mechanism (right-panel).
As expected, the observed physical parameters show significant deviations from these predicted values.

A pressing question arises: How unique is the FN charge assignment in replicating the observed parameters? 
In fact, multiple proposals in the literature suggest alternate ``optimal" FN charge assignments that seem to align with observations\,\cite{Fedele:2020fvh,Cornella:2023zme}.
Another question is that which charge assignment is ``better" among several charge assignments
when they seemingly fit the observations.
Besides, 
it is also interesting to consider whether 
there are ``good" FN charge 
assignments which are
consistent with the GUT.
In order to answer these questions, 
we systematically investigate the FN charges by using Bayesian inference, while taking into account the prior distribution of the $\order{1}$ coefficients.

\subsection{Some remarks}
\label{sec:remarks_FNmechanism}

There are several noteworthy considerations concerning the charge determination and the associated physical parameters in the FN mechanism. 
We will outline these remarks below and comment on the neutrino mass ordering.

\subsubsection{Why Gaussian?}

Note that our assumption that the 
$\order{1}$ coefficients $\kappa$ in Eqs.\,\eqref{eq:distribution_1} and \eqref{eq:distribution_2} have 
normal distributions 
leads to the following key properties:
\begin{itemize}
    \item The distribution's mean is 0, ensuring that no special point or direction is introduced in the flavor basis.
    \item The normal distribution  maximizes informational entropy, 
    $ - \int dx\, p(x) \log p(x)$, for given means and variances,
    where $p(x)$ is the probability density function of the continuous random variable $x$.
    This allows us to derive the least biased probability estimates.
    \item The common variance of $\sigma$ ensures a distribution without unnecessary bias within the flavor space. 
\end{itemize}
In the present work,
we set the common variance $\sigma$ to $1$ as a representative value
to realize $\kappa$ parameters with values that we intuitively perceive as $\order{1}$.

\subsubsection{Physical parameters}
\label{sec:Parameters}
In the above example, 
we compare the predictions with the running parameters 
at the renormalization scale
much higher energy than the electroweak scale.
This is because that 
in the FN mechanism, the effective coupling constants are generated at around the FN breaking scale,
which we assume to be at an extremely high energy scale for both the cutoff scale $M_*$ and the FN breaking scale $\langle\Phi\rangle$.

In our analysis,
we contrast the predictions of the FN mechanisms with the $\overline{\mathrm{MS}}$ parameters in the SM at a renormalization scale of 
$\mu_{R} = 10^{15}$\,GeV, 
which is assumed to represent the FN breaking scale.
It is also noted that the detailed choice of the renormalization scale does not have a significant impact as long as it is sufficiently higher than the electroweak scale when comparing the model's predictions with the running parameters.
This is because the QCD coupling and top Yukawa coupling become small at the high energy scale, 
so that the running effects become insignificant.
For example, 
the largest effect of 
the change of the 
renormalization scale from $\mu_R=10^{15}$\,GeV
to 
$\mu_R=10^{14}$\,GeV
is the change of the top Yukawa coupling by about $5$\%.
Such a small change can be easily absorbed by the $\order{1}$ distributions of the fundamental parameters $\kappa$'s.
For the neutrino parameters, on the other hand, we adopt low-energy observation values given by NuFit~5.2\,\cite{Esteban:2020cvm}, as discussed in Sec.\,\ref{sec:heuristic_example}.

\section{Bayesian Inference for Charge Assignment}
\label{sec:inference}
To quantitatively evaluate the fit quality of charge assignments, we conduct a Bayesian analysis and estimate the Bayes factor for each charge assignment. 
A similar approach has been utilized to compare various FN charge choices for the lepton sector in Ref.\,\cite{Bergstrom:2014owa}.

\subsection{Bayesian inference}
\label{sec:general_inf}

Let us consider a model $M_i$ labeled by $i$ with model parameters $\mathbf{\Theta}$.
In our case, the label of the model corresponds to the choice of the FN charge assignment as well as the neutrino mass models,
while $\mathbf{\Theta}$ includes the Lagrangian parameters such as $\kappa$ and $\epsilon$.
The prior probability of the model is denoted by $\mathrm{P}(M_i)$. 
The prior distribution of the model parameters
is denoted by $\mathrm{P}(\mathbf{\Theta}|M_i)$ which corresponds to the conditional probability density of $\mathbf{\Theta}$ for a given model $M_i$,
satisfying $\int d\mathbf{\Theta}\, \mathrm{P}(\mathbf{\Theta}|M_i)=1$.
The likelihood function $\mathrm{P}(\mathbf{D}|\mathbf{\Theta},M_i)$
with $\int d\mathbf{D}\, \mathrm{P}(\mathbf{D}|\mathbf{\Theta},M_i)=1$
is also the conditional probability distribution of
the experimental data $\mathbf{D}$ for given model parameters $\mathbf{\Theta}$ and model $M_i$.

With these distribution functions, 
the marginalized likelihood or the evidence of $M_i$ is defined as:
\begin{gather}
 \calZ_i
 = \mathrm{P}(\mathbf{D}|M_i)
  = \int \mathrm{P} (\mathbf{D}|\mathbf{\Theta}, M_i) \mathrm{P} (\mathbf{\Theta}|M_i)\, d\mathbf{\Theta} \ .
\end{gather}
Using the marginalized likelihood, 
the posterior probability of $M_{i}$ for given data $\mathbf{D}$ is obtained as:
\begin{align}
 \mathrm{P}(M_{i}|\mathbf{D}) 
 = \frac{\mathrm{P}(\mathbf{D}|M_{i}) \mathrm{P}(M_{i})}{\mathrm{P}(\mathbf{D})} 
 = \frac{\calZ_i \, \mathrm{P}(M_{i})}{\mathrm{P}(\mathbf{D})} \ .
\end{align}
Here, the normalization factor is $\mathrm{P}(\mathbf{D}) = \sum_{i} \mathrm{P}(\mathbf{D}|M_{i}) \mathrm{P}(M_{i})$, 
which corresponds to the total probability of obtaining experimental data.

Comparison of the goodness of fit
between models 
for given experimental data  $\mathbf{D}$
is possible by the ratio or the odds of  the posterior probabilities. 
That is, 
the larger the odds
\begin{align}
 \frac{\mathrm{P}(M_{i}|\mathbf{D})}{\mathrm{P}(M_{j}|\mathbf{D})} 
 = \frac{\calZ_i}{\calZ_j}  \frac{\mathrm{P}(M_{i})}{\mathrm{P}(M_{j})}\ ,
\end{align}
becomes, 
the more probable model $M_i$ is compared to $M_j$.
However, 
since we do not have any prior knowledge of $\mathrm{P}(M_i)$,  it is customary to assume that the prior probabilities of all models are equal.
In this case, the odds of posterior probabilities are given by the Bayes factor $B_{ij} = \mathcal{Z}_i/\mathcal{Z}_j $.
To interpret the Bayes factor, 
we use
the Jeffreys scale given in Tab.\,\ref{tab:jeffreys} (see Refs.\,\cite{Jeffreys:1939xee,Kass:1995loi}).

\begin{table}[t!]
\caption{This table shows the Jeffreys scale for assessing the strength of evidence in favor of one model over another. 
Under the assumption that both models have equal prior probabilities, the posterior probability of the preferred model is determined.}
  \label{tab:jeffreys}
 \begin{center}
  \begin{tabular}{c|c|c} \hline
    $\mathrm{log}_{10} (B_{ij})$ & $B_{ij}$ & Strength of evidence \\ \hline
    0 to 1/2 & 1 to 3.2 & Not worth more than a bare mention \\ 
    1/2 to 1 & 3.2 to 10 & Substantial \\
    1 to 2 & 10 to 100 & Strong \\
    $ >2 $ & $ >100 $ & Decisive \\ \hline
  \end{tabular}
 \end{center} 
\end{table}

\subsection{Prior distributions and likelihood function}
\label{sec:prior}

For $\kappa^{(u,d,e,D)}$, 
we adopt the prior distributions of $3 \times 3$ matrices, $\kappa$'s as,
\begin{align}
\pi(\kappa)
:=
\prod_{i,j}
\left(\frac{1}{\sqrt{2\pi}\sigma} \right)^2
\exp\left({{- \frac{|\kappa_{ij}|^{2}}{2\sigma^2}}}\right)
= 
\left(\frac{1}{\sqrt{2\pi}\sigma} \right)^{18} 
\exp\left({{- \frac{\tr{\kappa^\dagger\kappa}}{2\sigma^2}}}\right) \ .
\end{align}
For $\kappa^{(R,W)}$, 
\begin{align}
\pi(\kappa)
&:=
\qty[\prod_{i}
\left(\frac{1}{\sqrt{2\pi}  \sigma} \right)^2
\exp\left({{- \frac{|\kappa_{ii}|^{2}}{2 \sigma^2}}}\right)]
\times
\qty[\prod_{i>j}
\left(\frac{1}{\sqrt{\pi} \sigma} \right)^2
\exp \qty(-\frac{|\kappa_{ij}|^{2}}{\sigma^2} )] \nonumber \\
&= 
\left(\frac{1}{\sqrt{2\pi} \sigma} \right)^6
\left(\frac{1}{\sqrt{\pi} \sigma} \right)^6
\exp\left({{- \frac{\tr{\kappa^\dagger\kappa}}{2 \sigma^2}}}\right) \ .
\end{align}
The integral measure is
\begin{align}
\label{eq}
d\kappa := \prod_{i,j} 
d\, (\Re\,\kappa_{ij})\, 
d\,(\Im\, \kappa_{ij}) \ .
\end{align}
This prior distribution matches the random variables in Eqs.\,\eqref{eq:distribution_1} and \eqref{eq:distribution_2}.
Note that both the distributions and the integral measures of $\kappa$'s do not have specific flavor dependencies, motivated by our assumption that the flavor structure of the SM is purely governed by the FN charge assignments.
Accordingly, $\pi(\kappa)$ and $d\kappa$'s are invariant under $\mathrm{U}(3)$ flavor rotations.

Altogether, 
we adopt the prior distribution of the model parameters as follows,
\begin{align}
\mathrm{P}(\kappa,\epsilon|M_F) \, d\mathbf{\Theta}
= \frac{1}{\epsilon_{\max}-\epsilon_{\min}}
\, \prod_{i \in F}\pi(\kappa^{(i)}) \, 
d\kappa^{(i)}\, d\epsilon\ ,
\end{align}
where $\epsilon$ ranges in 
$\epsilon = [\epsilon_{\min},\epsilon_{\max}]$.
For the quark sector analysis, $i$ takes 
\begin{align}
   F= F_\mathrm{quark}=\{u,d\} \ .
\end{align}
For the lepton sector analysis,
$i$ takes 
\begin{align}
  F= F_\mathrm{lepton,\,seesaw}
  =\{e,D,R\} \ ,
\end{align}
for the seesaw mechanism, and 
\begin{align}
   F=F_\mathrm{lepton,\,\dimfive}
   =\{e,W\} \ ,
\end{align}
for the dimension-five operator.

For the likelihood function for the physical parameters
in App.\,\ref{sec:msbar},
$\mathbf{x}=\{x_i\}$,
we use
\begin{align}
\label{eq:lnL}
 \mathcal{L}(\mathbf{x}) 
 = 
 \prod_{i} \frac{1}{\sqrt{2\pi} \sigma_i}\exp\left( -\frac{ (x_i - {x}^\mathrm{obs}_i )^2 }{2 \sigma_i^2} \right) \simeq   \prod_{i} \delta (x_i - {x}^\mathrm{obs}_i ) \ .
\end{align}
Here, ${x}^\mathrm{obs}_i$ and $\sigma_i$ are the observed values and their errors, respectively.
In models that reasonably explain the observed values, 
the spread of the prior distribution is sufficiently larger than the observational errors,
the likelihood function can be approximated by the $\delta$-function, 
and hence we use this approximation except for $\delta_{\mathrm{13}}^\mathrm{PMNS}$ and $\Delta m_{31}^{2}$.

Before closing this section, 
let us give the Bayes factors
for the SM without FN mechanism.
In this case, the integration of the Bayes factor separates into the integration over the unitary matrices 
and the integration over the eigenvalues (see the App.\,\ref{sec:measure_yukawa}).
Thus,
the Bayes factor is given by the product of these separated contributions in this case.
For the quark sector,
\begin{align}
\calZ_{0,\mathrm{up\, quark}}
&= \frac{1}{2^8}\, y_u\, y_c\,
   y_t\,
   \qty(y_u^2-y_c^2)^2
   \qty(y_c^2-y_t^2)^2
   \qty(y_t^2-y_u^2)^2\,
   \mathrm{exp}
   \qty(-\frac{y_u^2+y_c^2+y_t^
   2}{2})
= 3.4 \times 10^{-26}\ , \\
\calZ_{0,\mathrm{down\, quark}}
&= \frac{1}{2^8}\, y_d\, y_s\,
   y_b\,
   \qty(y_d^2-y_s^2)^2
   \qty(y_s^2-y_b^2)^2
   \qty(y_b^2-y_d^2)^2\,
   \mathrm{exp}
   \qty(-\frac{y_d^2+y_s^2+y_b^
   2}{2})
= 8.0 \times 10^{-48}\ , \\
\calZ_{0,\mathrm{CKM}}
&= \frac{8}{\pi}\, (1-\sin^2 \theta^{\mathrm{CKM}}_{13}) \times 
\sin \theta^{\mathrm{CKM}}_{12} \times
\sin \theta^{\mathrm{CKM}}_{23} \times 
\sin \theta^{\mathrm{CKM}}_{13} 
= 1.1 \times 10^{-4}\ .
\end{align}
Here, 
we have used the SM parameters in the App.\,\ref{sec:msbar}.
Similarly, for the lepton sector with the dimension-five operators and the seesaw mechanism, 
\begin{gather}
\calZ_{0,\mathrm{charged\, lepton}}
= \frac{1}{2^8}\, y_e\, y_{\mu}\,
   y_{\tau}\,
   \qty(y_e^2-y_{\mu}^2)^2
   \qty(y_{\mu}^2-y_{\tau}^2)^2
   \qty(y_{\tau}^2-y_e^2)^2\,
   \mathrm{exp}
   \qty(-\frac{y_e^2+y_{\mu}^2+y_{\tau}^
   2}{2})
= 5.0 \times 10^{-43}\ , \\
\calZ^{\mathrm{NO}}_{0,\dimfive,\nu\mathchar`-\mathrm{mass}}
= 
\frac{32 (1-r) r}{(r+1)^5}
= 0.79\ ,  \quad
\calZ^{\mathrm{NO}}_{0,\seesaw,\nu\mathchar`-\mathrm{mass}}
= 6.7\ , \\
\calZ_{0,\mathrm{PMNS}}
= \frac{1}{\pi}\, (1-\sin^2 \theta^{\mathrm{PMNS}}_{13})
= 0.31\ ,
\label{eq:ZPMNS}
\end{gather}
where $r = \Delta m^2_{12}/\Delta m^2_{13}$.
Therefore,
\begin{gather}
\calZ_{0, \mathrm{quark}}
= \calZ_{0, \mathrm{up\,quark}} \times \calZ_{0, \mathrm{down\,quark}} \times \calZ_{0, \CKM}
= 3.0 \times 10^{-77}\ , \\
\calZ^{\mathrm{NO}}_{0, \dimfive}
= \calZ_{0,\mathrm{charged\, lepton}} \times \calZ^{\mathrm{NO}}_{0,\dimfive,\nu\mathchar`-\mathrm{mass}}
\times \calZ_{0,\mathrm{PMNS}}
= 1.2 \times 10^{-43}\ , \\
\calZ^{\mathrm{NO}}_{0, \mathrm{\seesaw}}
= \calZ_{0,\mathrm{charged\, lepton}} \times \calZ^{\mathrm{NO}}_{0,\seesaw,\nu\mathchar`-\mathrm{mass}}
\times \calZ_{0,\mathrm{PMNS}}
= 1.0 \times 10^{-42}\ .
\end{gather}
Due to the small Yukawa couplings of the first and second generations,
the Bayes factor is significantly suppressed in the SM without FN mechanism.

\section{Comparison of FN Charge Assignment}
\label{sec:result_charge}

In this section, 
we present the results of analyzing the FN charge assignments for the following cases:
\begin{itemize}
    \item Only the quark sector,
    \item Only the lepton sector with the seesaw mechanism or the dimension-five operators,
    \item Both the quark and lepton sectors.
\end{itemize}
We also discuss the compatibility of FN charge assignments motivated by the SU(5) GUT multiplets.

\subsection{Search strategy of FN charges}
\label{sec:charge_choice}

In this work, 
we consider the range of the FN charges to be $|f_{Q,\ubar,\dbar,L,\ebar}|\le 10$ and the range of $\epsilon$ to be $\epsilon \in [0.05, 0.35]$.
In addition,
for the seesaw mechanism,
we assume all $f_{\Nbar,a}=0$ ($a=1,2,3$).
In this parameter space,
we eliminate the degrees of freedom associated with the redundancy of FN charge assignments,
which result in the same structure of the Yukawa matrices.
The examples of redundancy are permutation of generations,
overall sign
and possible constant shift of FN charges which do not affect the flavor structure,%
\footnote{For instance, 
a charge assignment $f_L=(5,4,4)$ and $f_{\ebar}=(5,2,0)$ is equivalent to 
an assignment $f_L=(3,2,2)$ and $f_{\bar{e}}=(7,4,2)$.
It is sufficient to consider the Bayes factor for any one of the charge assignments.}
etc.
By reducing redundancies,
we consider about $4 \times 10^{8}$ charges for quark sector,
and about $10^{6}$ charges for lepton sector.

As experimental data, 
we use the Yukawa couplings and (CKM/PMNS) mixing parameters in App.\,\ref{sec:msbar}.
Note that, 
in our analysis,
we use fit results with Super-Kamiokande (SK) data, and we mainly focus on the normal ordering case.
Examples of FN charges consistent with inverted ordering and the corresponding Bayes factors are provided in the App.\,\ref{sec:inverted_comment}.
We use Monte-Carlo integration to calculate the Bayes factor, 
incorporating nested sampling\,\cite{Skilling:2004pqw} as part of the algorithm. 
These results are cross-checked using UltraNest\,\cite{Buchner_2019,UltraNest2,UltraNest3}.

\subsection{Quark sector}
\label{sec:quark_result}

In Tab.\,\ref{tab:qres03}, 
examples of the FN charges with large Bayes factors are listed.
The $f_{Q,\ubar,\dbar}$ are arranged in generational order. 
This order is chosen so that the flavor eigenstates approximately correspond to the mass eigenstates.
The range of $\epsilon$ represents the 95\% credible interval (CI) of its posterior distribution.
The results show that the models with charge assignments in Tab.\,\ref{tab:qres03}
have significantly larger Bayes factors compared to models
without FN charges,
highlighting the success of the FN mechanism (see also Tab.\,\ref{tab:jeffreys}).

\begin{table}[t!]
\caption{Ten FN charge assignments of quarks with large Bayes factor.
These FN charges are arranged from the first generation to the third generation from left to right.
$\calZ_{\quark}$ is the marginalized likelihoods for the cases where quarks have the FN charges shown in the table.
The range of $\epsilon$ is the 95\% CI of its posterior distribution.
}
 \begin{center}
  \begin{tabular}{|c|c|c|c|c|} \hline
    $f_Q$ & $f_{\ubar}$ & $f_{\dbar}$ & $\mathrm{log}_{10} (\calZ_{\quark}/\calZ_{0,\quark})$ & $\epsilon$ \\ \hline
    $5,3,0$ & $7,3,1$ & $6,5,5$ & $97.34 \pm 0.02$ & $ 0.302 \to 0.326 $ \\
    $5,3,0$ & $7,3,1$ & $7,5,5$ & $97.16 \pm 0.05$ & $ 0.310 \to 0.333 $ \\
    $3,2,0$ & $4,2,0$ & $3,3,3$ & $97.03 \pm 0.03$ & $ 0.126 \to 0.145 $ \\ 
    $5,3,0$ & $7,4,1$ & $6,5,5$ & $96.89 \pm 0.02$ & $ 0.311 \to 0.335 $ \\ 
    $6,4,0$ & $7,3,-1$ & $5,5,5$ & $96.83 \pm 0.13$ & $ 0.316 \to 0.338 $ \\
    $3,2,0$ & $6,3,1$ & $5,4,4$ & $96.64 \pm 0.03$ & $ 0.207 \to 0.230 $ \\
    $4,3,0$ & $6,3,0$ & $5,4,4$ & $96.60 \pm 0.04$ & $ 0.238 \to 0.261 $ \\
    $6,4,0$ & $6,3,-1$ & $5,5,4$ & $96.48 \pm 0.05$ & $0.299 \to 0.322$ \\
    $6,4,0$ & $6,3,-1$ & $5,5,5$ & $96.42 \pm 0.02$ & $0.312 \to 0.336$ \\ 
    $4,4,0$ & $4,2,-1$ & $3,-10,4$ & $95.74 \pm 0.26$ & $0.186 \to 0.209$ \\  \hline
  \end{tabular}
  \label{tab:qres03}
 \end{center} 
\end{table}

\begin{figure}[t!]
	\centering
 	\subcaptionbox{}
	{\includegraphics[width=0.49\textwidth]{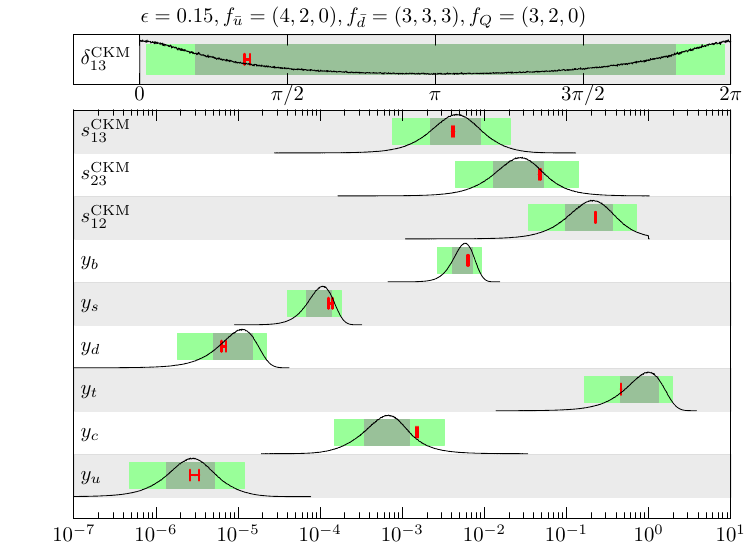}}
 	\subcaptionbox{}
	{\includegraphics[width=0.49\textwidth]{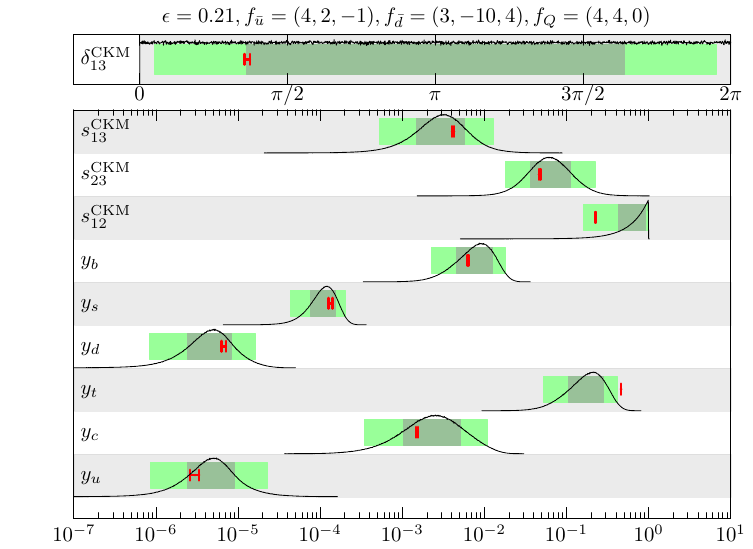}}
 \caption{Predictions from $\order{1}$ distributions of $\kappa$'s with the FN charge assignments in Tab.\,\ref{tab:qres03}.
The meanings of the red bars and band colors are the same as in Fig.\,\ref{fig:conv}.
 }
 \label{fig:best_quark}
\end{figure}

This is further confirmed by the distribution of the predicted physical parameters for the FN charges listed in Tab.\,\ref{tab:qres03}.
Fig.\,\ref{fig:best_quark} shows that the observed physical parameters fall within the
typical range of the predictions of the FN mechanism with the FN charges in Tab.\,\ref{tab:qres03}.
The light green bands, 
dark green bands and red bars have the same meanings as before 
(see Sec.\,\ref{sec:heuristic_example}).

\begin{figure}[t!]
	\centering
{\includegraphics[width=0.90\textwidth]{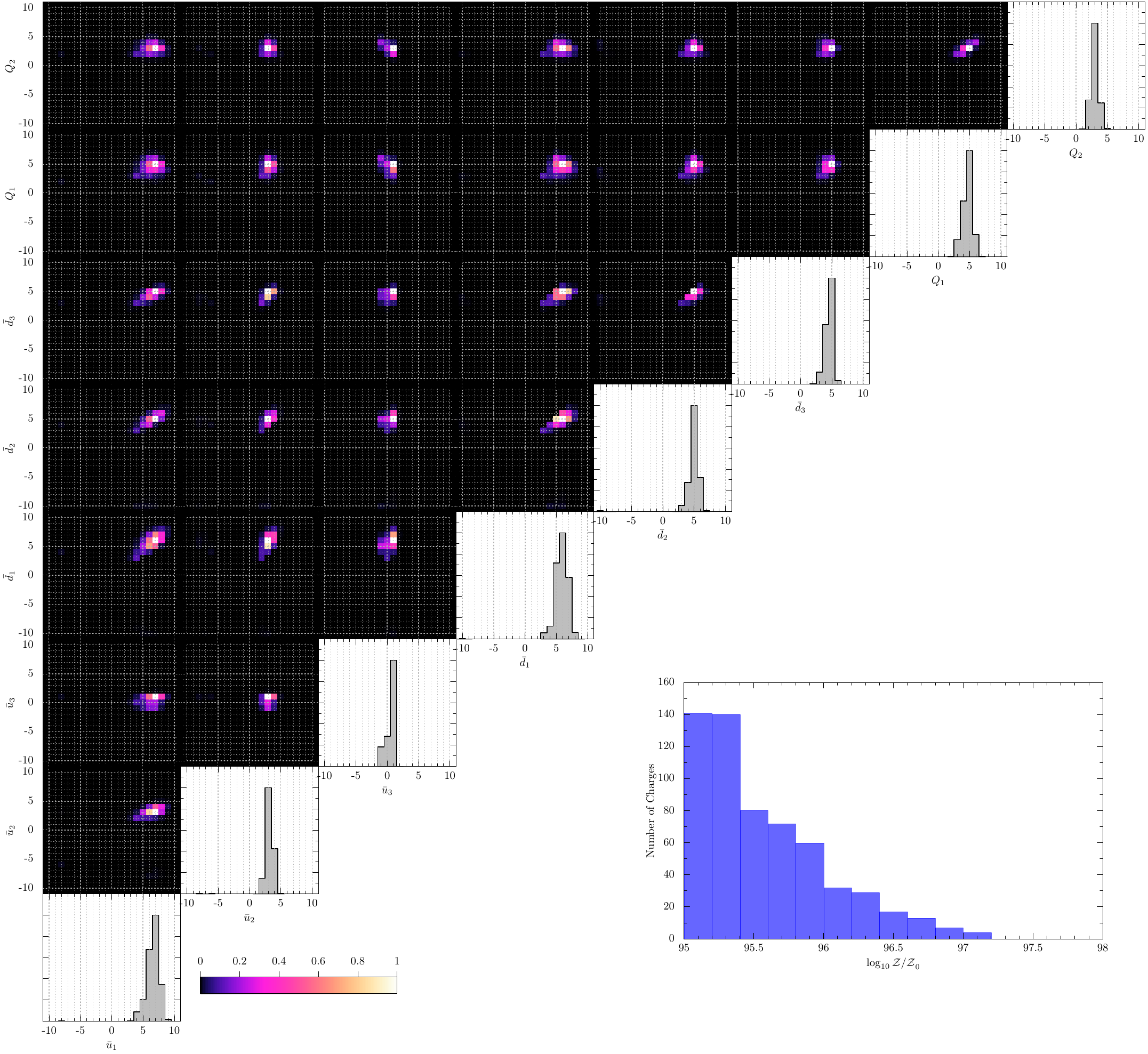}}
 \caption{The upper-left panel shows a two-dimensional histogram of
 the quark charges.
 These results are normalized after being weighted by the values of the Bayes factor.
 The lower-right panel shows the distribution of Bayes factors of the quark FN charge candidates.
 }
 \label{fig:quark_corner}
\end{figure}

The FN charge assignments in Tab.\,\ref{tab:qres03} are notable in two points.
First, 
some FN charge assignments correspond to $q^{(u)}_{33}=1$. 
In the most previous studies,
the FN charges of $Q_3$ and $\ubar_3$
are assumed to be $0$ 
so that $q^{(u)}_{33}=0$
because the top Yukawa coupling is of $\order{1}$. 
However, 
considering the effects of the renormalization group, 
the top Yukawa coupling above the electroweak scale is approximately half of its low-energy value (see the App.\,\ref{sec:msbar}).
Second, 
there is the FN charge assignment where 
some of the FN charges
have negative signs,
such as $f_Q = (6,4,0)$, 
$f_{\ubar}=(7,3,-1)$, 
$f_{\dbar}=(5,5,5)$ or $f_Q = (4,4,0)$, 
$f_{\ubar}=(4,2,-1)$, 
$f_{\dbar}=(3,-10,4)$,
which can successfully reproduce the flavor structure.
The discovery of such negative charges was made possible by setting a broad search range for the charges, $ |f_{Q,\ubar,\dbar}| \leq 10 $,
which are not considered in other literature\,\cite{Cornella:2023zme}. 
Note also that some of our favorable FN charges overlap with the FN charges discussed in Refs.\,\cite{Fedele:2020fvh,Cornella:2023zme} up to redundancy of FN charge assignments.%
\footnote{ 
The FN charge assignment with $\epsilon \rightarrow \epsilon^{1/N}$ and $f \rightarrow N \times f$ is equivalent to the original assignment with $\epsilon$ and $f$.}

Fig.\,\ref{fig:quark_corner} shows a two-dimensional histogram of the FN charges, 
which are normalized after being weighted by the values of the Bayes factor.
In this figure, 
we resolve the charge redundancy 
by setting  $f_{Q,3}=0$ and imposing $f_{Q,1} \ge 0$.
In the figure, we also show the distribution of Bayes factors.
The figure demonstrates that there are $\order{10^3}$ FN charge assignments capable of reproducing the flavor structure of the SM.

\subsection{Lepton sector}
\label{sec:lepton_result}

Tabs.\,\ref{tab:sres03} and \ref{tab:wres03} present examples of charge assignments with large Bayes factors. 
In particular, 
the top charge assignment in each table corresponds to the one with the largest Bayes factor for the respective case.
The range of $\epsilon$ represents the 95\% CI of its distribution.
The results show that the models with charge assignments in Tabs.\,\ref{tab:sres03} and \ref{tab:wres03}
have significantly larger Bayes factors compared to models
without FN charges,
highlighting the success of the FN mechanism (see also Tab.\,\ref{tab:jeffreys}).

\begin{figure}[t!]
    \centering
    \begin{subfigure}[b]{0.49\textwidth}
        \centering
        \includegraphics[width=\textwidth]{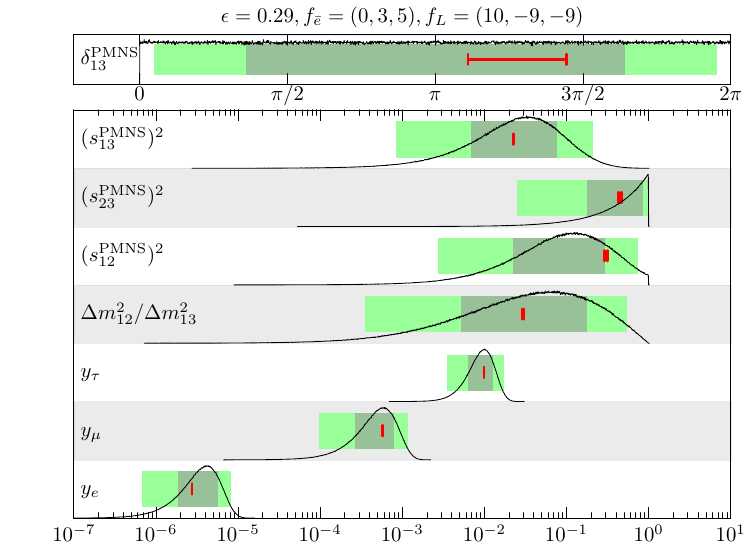}
        \caption{}
        \label{fig:seesaw1}
    \end{subfigure}
    \hfill
    \begin{subfigure}[b]{0.49\textwidth}
        \centering
        \includegraphics[width=\textwidth]{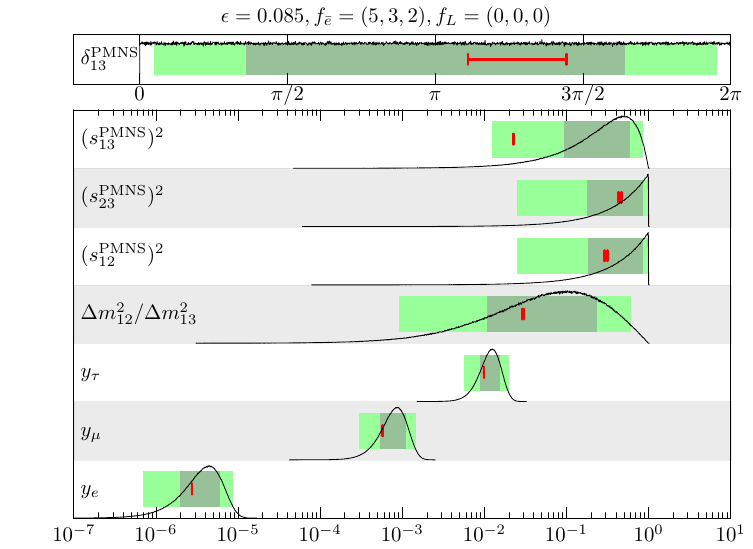}
        \caption{}
        \label{fig:seesaw2}
    \end{subfigure}
    \vskip\baselineskip
    \begin{subfigure}[b]{0.49\textwidth}
        \centering
        \includegraphics[width=\textwidth]{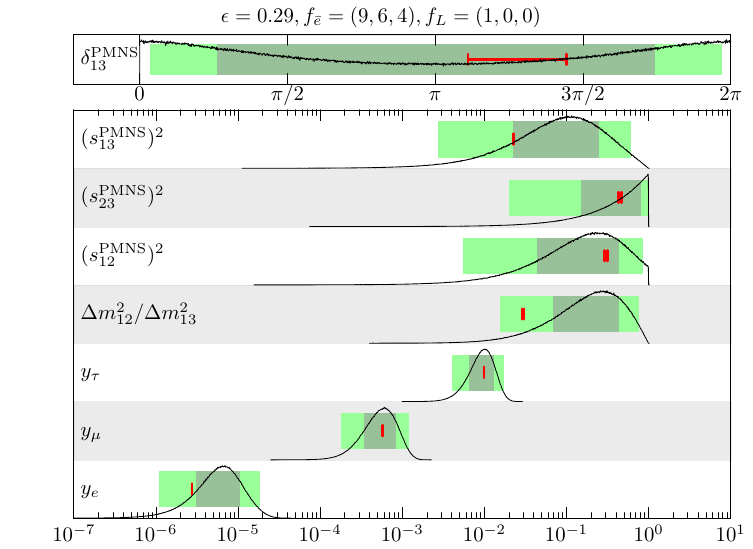}
        \caption{}
    \end{subfigure}
    \hfill
    \begin{subfigure}[b]{0.49\textwidth}
        \centering
        \includegraphics[width=\textwidth]{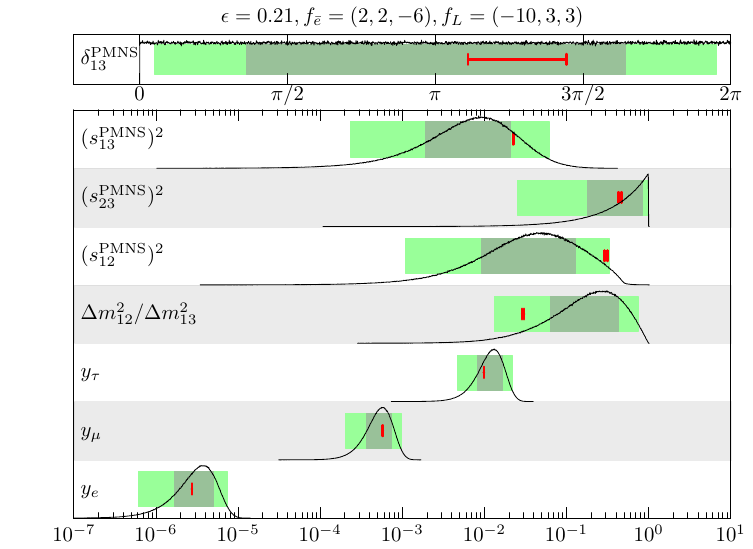}
        \caption{}
    \end{subfigure}
    
\caption{Figure (a) and (b) show 
predictions from $\order{1}$ distributions of $\kappa$'s with the FN charge assignments in Tab.\,\ref{tab:sres03} in the seesaw mechanism.
Figure (c) and (d) show
predictions from $\order{1}$ distributions of $\kappa$'s with the FN charge assignments in Tab.\,\ref{tab:wres03} in the dimension-five operator case.
The meanings of the red bars and band colors are the same as in Fig.\,\ref{fig:conv}.}
  \label{fig:lepton_compare}
\end{figure}

This is further confirmed by the distribution of the predicted physical parameters for the FN charges listed in Tabs.\,\ref{tab:sres03} and \ref{tab:wres03}.
Fig.\,\ref{fig:lepton_compare} shows that the observed physical parameters fall within the
typical range of the predictions of the FN mechanism with the FN charges in Tabs.\,\ref{tab:sres03} and \ref{tab:wres03}.
The light green bands, 
dark green bands and red bars have the same meanings as before 
(see Sec.\,\ref{sec:heuristic_example}).

\begin{table}[t!]
\caption{For the lepton sector with the seesaw mechanism. 
Ten FN charge assignments of leptons with large Bayes factors.
These FN charges are arranged from the first generation to the third generation from left to right.
$\calZ_{\seesaw}$ is the marginalized likelihoods for the cases where leptons have the FN charges shown in the table.
The range of $\epsilon$ is the 95\% CI of its posterior distribution.}
 \begin{center}
  \begin{tabular}{|c|c|c|c|} \hline
    $f_L$ & $f_{\ebar}$ & $\mathrm{log}_{10} (\calZ_{\seesaw}/\calZ_{0,\seesaw})$ & $\epsilon$ \\ \hline
    $9,-8,-8$ & $1,2,4$ & $52.59 \pm 0.03$ & $0.264 \to 0.309$ \\
    $10,-9,-9$ & $0,3,5$ & $52.58 \pm 0.07$ & $0.264 \to 0.308$ \\ 
    $2,-1,-1$ & $8,7,5$ & $52.42 \pm 0.13$ & $0.258 \to 0.292$ \\
    $5,4,4$ & $5,2,0$ & $52.29 \pm 0.04$ & $0.261 \to 0.298$ \\
    $2,1,-1$ & $8,5,5$ & $52.27 \pm 0.09$ & $0.257 \to 0.294$ \\
    $2,-1,-1$ & $6,6,4$ & $52.26 \pm 0.10$ & $0.186 \to 0.224$ \\
    $5,4,4$ & $6,2,0$ & $52.19 \pm 0.05$ & $0.278 \to 0.313$ \\
    $8,0,0$ & $10,6,-4$ & $52.13 \pm 0.10$ & $0.271 \to 0.302$ \\
    $5,5,4$ & $6,2,0$ & $52.07 \pm 0.07$ & $0.295 \to 0.327$ \\
    $4,3,3$ & $5,2,0$ & $52.01 \pm 0.08$ & $0.209 \to 0.234$ \\ \hline 
  \end{tabular}
  \label{tab:sres03}
 \end{center} 
\end{table}
\begin{table}[t!]
\caption{For the lepton sector with the dimension-five operator. 
Ten FN charge assignments of leptons with larger logarithms of Bayes factors with respect to the case where all leptons have no FN charge.
These FN charges are arranged from the first generation to the third generation from left to right.
$\calZ_{\dimfive}$ is the marginalized likelihoods for the cases where leptons have the FN charges shown in the table.
The range of $\epsilon$ represents the 95\% CI of its posterior distribution.}
 \begin{center}
  \begin{tabular}{|c|c|c|c|} \hline
    $f_L$ & $f_{\ebar}$ & $\mathrm{log}_{10} (\calZ_{\dimfive}/\calZ_{0,\dimfive})$ & $\epsilon$ \\ \hline
    $2,0,-1$ & $10, 7, 5$ & $53.78 \pm 0.03$ & $0.311 \to 0.342$ \\
    $-10,3,3$ & $2,2,-6$ & $53.25 \pm 0.03$ & $0.188 \to 0.221$ \\
    $-10,3,3$ & $2,2,0$ & $53.22 \pm 0.03$ & $0.187 \to 0.222$ \\
    $5,5,4$ & $6,2,0$ & $53.19 \pm 0.03$ & $0.294 \to 0.334$ \\
    $5,4,4$ & $5,2,0$ & $53.14 \pm 0.03$ & $0.258 \to 0.294$ \\
    $4,-1,-1$ & $7,7,5$ & $53.14 \pm 0.03$ & $0.280 \to 0.317$ \\
    $1,0,0$ & $-9,5,3$ & $53.04 \pm 0.12$ & $0.188 \to 0.220$ \\
    $4,4,3$ & $5,2,0$ & $53.00 \pm 0.03$ & $0.223 \to 0.260$ \\
    $10,1,0$ & $10,7,-5$ & $52.96 \pm 0.03$ & $0.316 \to 0.343$ \\ 
    $4,3,3$ & $5,2,0$ & $52.68 \pm 0.04$ & $0.205 \to 0.235$ \\ 
    $2,1,0$ & $8,6,3$ & $52.15 \pm 0.03$ & $0.266 \to 0.301$ \\
    $0,0,0$ & $5,3,2$ & $51.49 \pm 0.04$ & $0.069 \to 0.083$ \\
    $1,0,0$ & $6,5,3$ & $51.48 \pm 0.05$ & $0.181 \to 0.211$ \\ 
    \hline
  \end{tabular}
\label{tab:wres03}
 \end{center} 
\end{table}

\begin{figure}[t!]
	\centering
	{\includegraphics[width=0.75\textwidth]{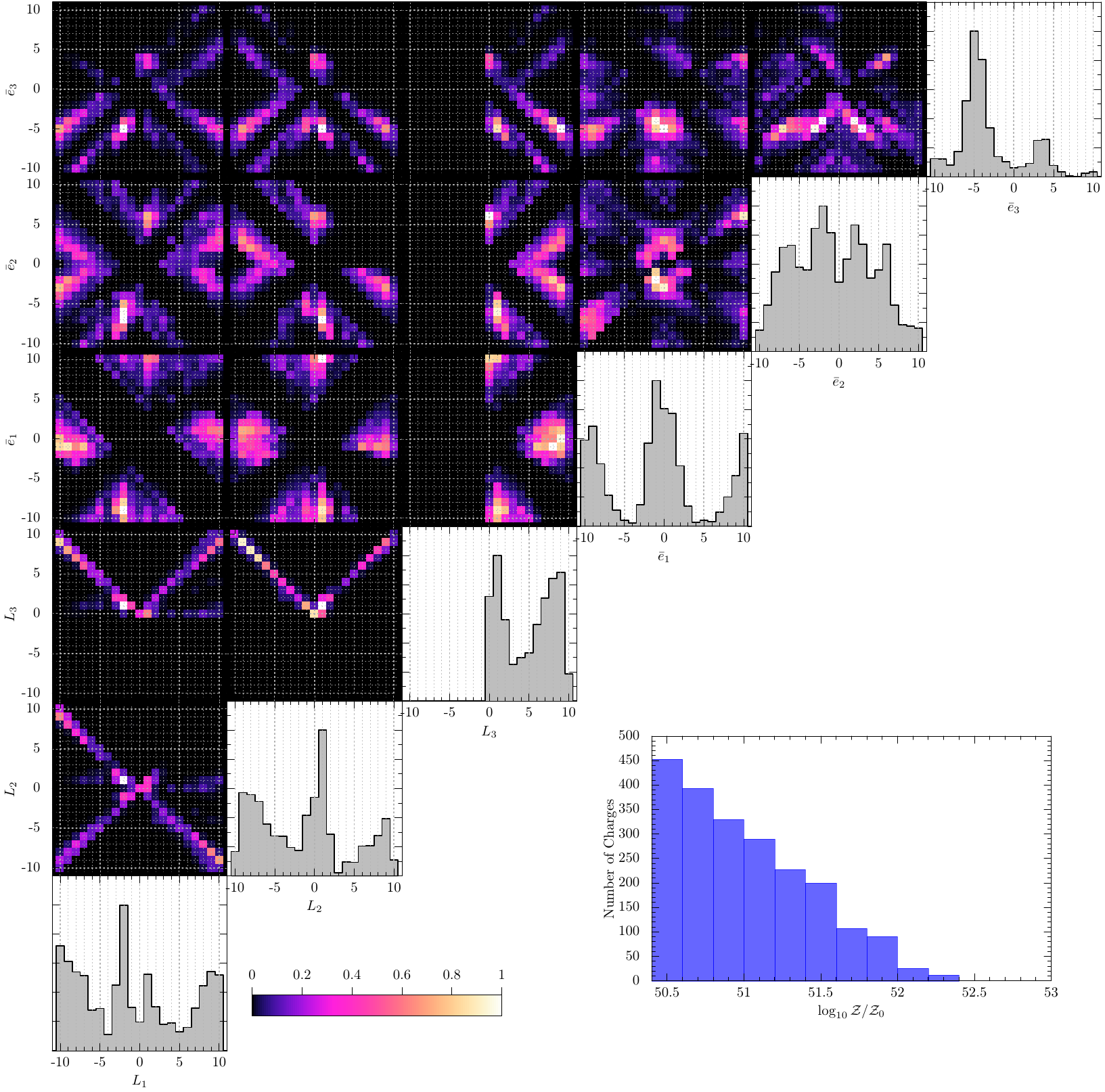}}
 \caption{The upper-left panel shows a two-dimensional histogram of our the lepton charges in the seesaw mechanism. 
 These results are normalized after being weighted by the values of the Bayes factor.
 The lower-right panel shows the distribution of Bayes factors of the lepton FN charge candidates.}
 \label{fig:seesaw_corner}
\end{figure}

\begin{figure}[t]
	\centering
	{\includegraphics[width=0.75\textwidth]{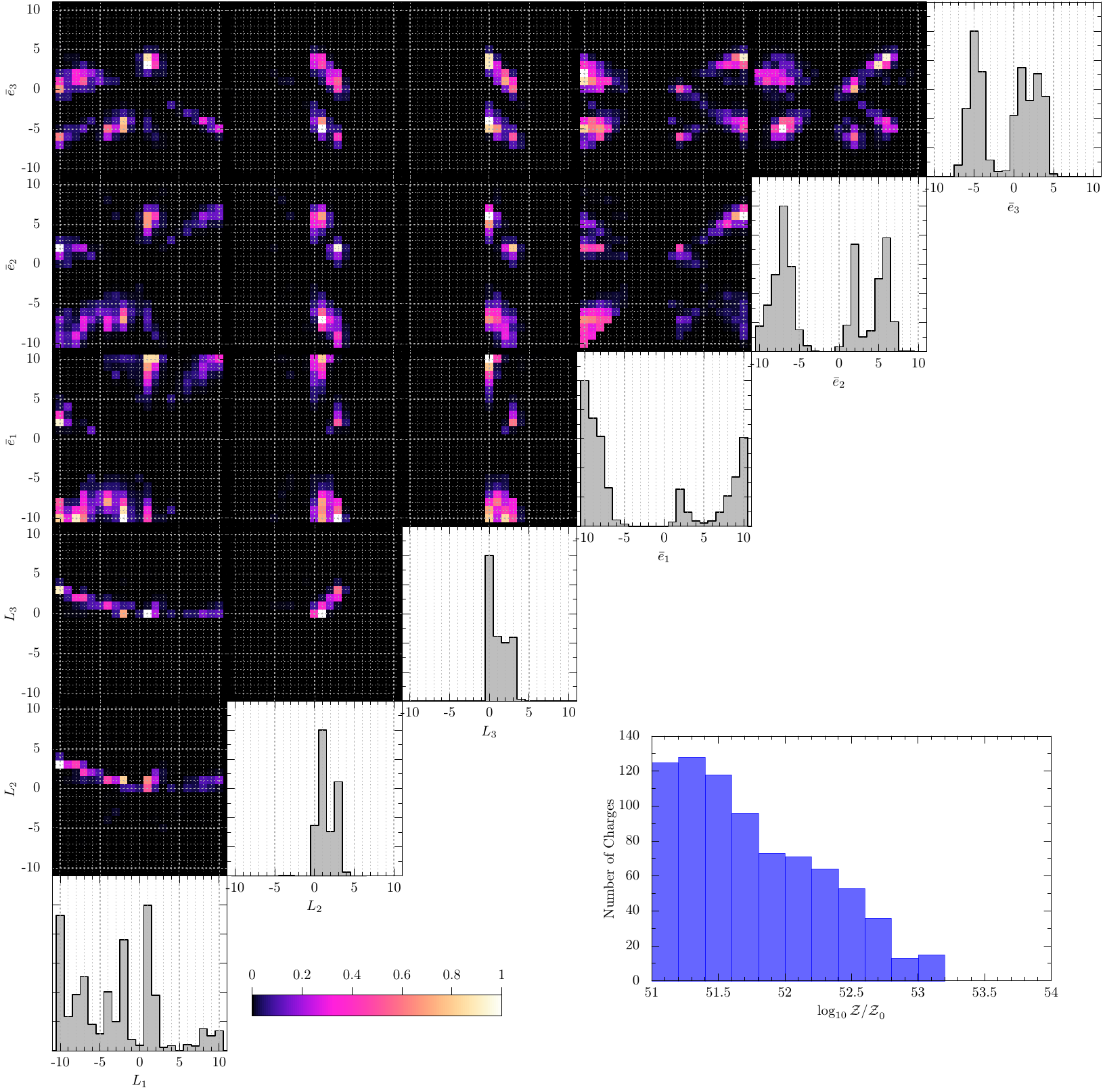}}
 \caption{The upper-left panel shows a two-dimensional histogram of lepton charges in the dimension-five operator case. 
 These results are normalized after being weighted by the values of the Bayes factor.
 The lower-right panel shows the distribution of Bayes factors of the lepton FN charge candidates.}
 \label{fig:dim5_corner}
\end{figure}

The FN charges in Tabs.\,\ref{tab:sres03} and \ref{tab:wres03} are notable for two reasons.
First, 
charge assignments including both positive and negative values are frequently observed in both the seesaw mechanism and dimension-five operator cases.
While previous studies assumed lepton charge assignments without negative values,
the results in Tabs.\,\ref{tab:sres03} and \ref{tab:wres03} demonstrate that negative charges can also yield sufficiently large Bayes factors and successfully reproduce the flavor structure.
Secondly, 
large generational differences in $f_L$ values are also permissible. 
Lepton charges that have been extensively studied (see Eq.\,\eqref{eq:charge_example}),
were characterized by minimal generational differences in $f_L$. 
However, 
our analysis demonstrates that even the FN charges with large generational differences, 
such as $f_L = (8, 0, 0)$ or $f_L = (-10, 3, 3)$, 
can successfully reproduce the flavor structure.

The three FN charges at the bottom of Tab.\,\ref{tab:wres03} correspond to those investigated in Ref.\,\cite{Bergstrom:2014owa}. 
They are referred to in Ref.\,\cite{Bergstrom:2014owa} as ``Hierarchy,'' ``New Anarchy,'' and ``$\mu \tau$-Anarchy,'' from top to bottom.  
However, 
whereas Ref.\,\cite{Bergstrom:2014owa} evaluates Bayes factors using fermion mass ratios as observables, 
we take the charged lepton Yukawa couplings themselves as observables. Consequently, 
the FN charge assignments presented in the table incorporate a flavor independent constant shift for $f_{\ebar}$.  
It is found that these charge assignments yield Bayes factors that are one to two orders of magnitude smaller than that of the best FN charge assignment. 
While Ref.\,\cite{Bergstrom:2014owa} examines five different charge assignments, 
it is shown that the types referred to as ``Anarchy'' and ``New Hierarchy'' exhibit sufficiently small Bayes factors compared to the best assignment.

Fig.\,\ref{fig:seesaw_corner}  and \ref{fig:dim5_corner} show a two-dimensional histogram of the FN charges, 
which are normalized after being weighted by the values of the Bayes factor for both cases.
In this figure, 
we resolve the charge redundancy 
by imposing $f_{L,3} \ge 0$.
In the figure, 
we also show the distribution of Bayes factors.
The figure demonstrates that there are $\order{10^3}$ FN charge assignments capable of reproducing the flavor structure of the SM as in the case of the quark charges.
Fig.\,\ref{fig:seesaw_corner} also exhibits a strong correlation among $f_L$ values across all generations. This indicates that even if $f_L$ is identical for all generations, 
it can still adequately account for the lepton flavor structure, 
as is also evident from Fig.\,\ref{fig:seesaw2} (upper-right panel).
Fig.\,\ref{fig:dim5_corner} also shows a strong correlation between $f_{L,2}$ and $f_{L,3}$ for $0\le f_{L,2}, f_{L,3}\le 4$, 
indicating that for charge assignments with large Bayes factors, 
the difference between $f_{L,2}$ and $f_{L,3}$ tends to be small. 
This suggests that $\sin^{2} \theta_{23}^{\PMNS}$ is well reproduced more frequently in such cases.

\subsubsection*{Comparison between seesaw mechanism and dimension-five operator}

Now, 
we discuss whether the dimension-five operator or the seesaw mechanism better explains the flavor structure.
For that purpose, 
let us compare the 
marginal likelihoods for each model,
\begin{align}
\frac{\calZ_{\seesaw}}{\calZ_{\dimfive}}
= 
\frac{\calZ_{\seesaw}}{\calZ_{0,\seesaw}}\,
\frac{\calZ_{0,\seesaw}}{\calZ_{0,\dimfive}}\,
\frac{\calZ_{0,\dimfive}}{\calZ_{\dimfive}}\ , 
\end{align}
where the Bayes factors  $\calZ_{\seesaw}/\calZ_{0,\seesaw}$
and $\calZ_{\dimfive}/\calZ_{0,\dimfive}$ are given in Tabs.\,\ref{tab:sres03} and \ref{tab:wres03}.

The ratio of the marginal likelihoods of the each model without the FN charges 
is given by,
\begin{align}
\label{eq:z0ratio}
\log_{10}\frac{\calZ_{0,\seesaw} }{\calZ_{0,\dimfive}}
= 0.92\ . 
\end{align}
The variation in the values of $\calZ_{0,\seesaw}$ and $\calZ_{0,\dimfive}$ stems from the differences in how the two models predict the neutrino mass difference ratio,
$(\Delta m_{12}^{2}/\Delta m_{13}^{2})$.
In the seesaw mechanism,
the predicted distribution for $(\Delta m_{12}^{2}/\Delta m_{13}^{2})$ is roughly an order of magnitude broader compared to that of the dimension-five operator.
This broader distribution enables the seesaw mechanism to accurately reproduce $(\Delta m_{12}^{2}/\Delta m_{13}^{2})$ even in the absence of FN charges. 
Consequently, 
as shown in Eq.\,\eqref{eq:z0ratio}, 
the absence of FN charges leads to a stronger preference for the seesaw mechanism over the dimension-five operator.

Comparing the FN charges with the largest Bayes factor in Tabs.\,\ref{tab:sres03} and \ref{tab:wres03}, 
we find that the marginal likelihoods are close with each other,
\begin{align}
\log_{10} \frac{\calZ_{\seesaw}}{\calZ_{\dimfive}} 
= 0.09\ .
\end{align}
This result indicates that 
under an appropriate FN charge assignment, the dimension-five operator has a more improved marginal likelihood compared to the seesaw mechanism.
Such an improvement is caused by a narrow prediction range on the mass difference ratio, $(\Delta m_{12}^{2}/\Delta m_{13}^{2})$ 
for the dimension-five operator compared with the seesaw mechanism
(see Fig.\,\ref{fig:lepton_compare}).
This result shows that there is no significant difference in preference between the seesaw mechanism and the dimension-five operator for the charge assignments with the largest Bayes factors in each case.

\subsection{Combined case}
\label{sec:combined_result}
Here, we analyze the Bayes factors 
of the charge assignments 
which include both the quark and lepton sectors simultaneously. 
In this case, the expression for the evidence can be obtained as follows: 
\begin{gather}
\label{eq:combined_evidence}
\calZ_{\mathrm{C}} 
=
\calZ_{\mathrm{quark}}\, 
\calZ_{l}\, 
\int d \epsilon \, 
\rho_q(\epsilon)\, \rho_l(\epsilon)\ ,
\quad  (l = \mathrm{\seesaw/\dimfive}) \ ,
\\  
\int d \epsilon\, \rho_{q,l}(\epsilon)=1\ .
\end{gather}
Here, $\calZ_\mathrm{quark}$ and $\calZ_l$ 
are the values of evidence of each sector.
The combined Bayes factors also depend 
on the posterior distributions of $\epsilon$, 
$\rho_q(\epsilon)$ ($\rho_l(\epsilon)$), 
for each sector
which have been obtained in the analyses 
of the each sector 
in the previous subsections.
The derivation of Eq.\,\eqref{eq:combined_evidence} is given in App.\,\ref{app:combined_evidence}. 

Note that we have fixed the sign convention of the FN charges in the quark and the lepton sector independently.
The relative sign of the FN charges between the two sectors, however,  does not affect the combined result, since the relevant flavor structures are independent in the quark and lepton sectors.

Tabs.\,\ref{tab:nongut_seesaw_res03} and \ref{tab:nongut_dim5_res03} list ten examples of FN charge assignments with large Bayes factors for the seesaw mechanism and dimension-five operators, respectively.
Here,
$\calZ_{\mathrm{C}}$ and $\calZ_{0,\mathrm{C}}$ denote the quark-lepton combined marginalized likelihoods 
for a given FN charge assignment and no FN charge, respectively.
The meaning of the range of 
$\epsilon$ is the same as before.
According to the Jeffreys scale in Tab.\,\ref{tab:jeffreys}, 
we have found that the all cases with FN charge assignments in Tabs.\,\ref{tab:nongut_seesaw_res03} and \ref{tab:nongut_dim5_res03} are strongly favored over the case with no FN charges.
Note that 
when the quark and lepton sectors are combined, 
the charge assignment with the largest Bayes factor is not necessarily a combination of the charges with the largest Bayes factors from each individual sector.
This is caused by the 
mismatch of the posterior distributions of $\epsilon$ for the quark and the lepton sectors in the separated analysis, 
which leads to
\begin{align}
    \int d \epsilon \, 
\rho_q(\epsilon)\, \rho_l(\epsilon) \ll 1 \ .
\end{align}

\begin{figure}[t!]
	\centering
	{\includegraphics[width=0.55\textwidth]{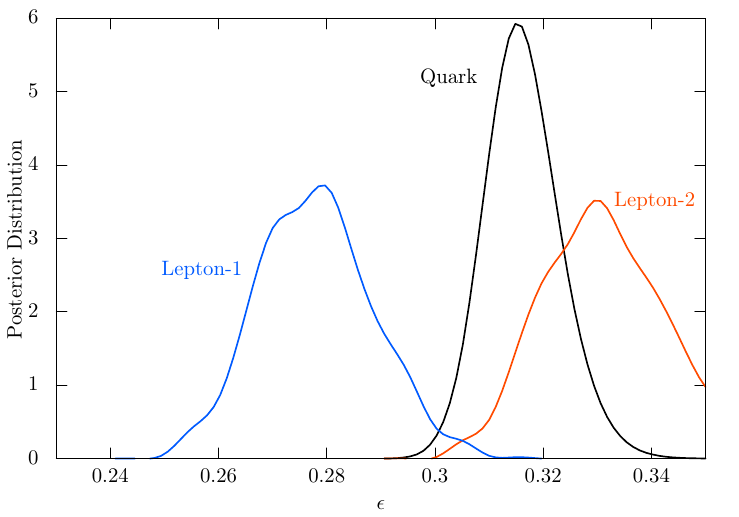}}
\caption{Posterior distributions of $\epsilon$ for the 
FN charge assignments in Eqs.\,\eqref{eq:quark}, \eqref{eq:lepton1} and \eqref{eq:lepton2} in the case of 
the dimension-five operator.}
 \label{fig:ep_distribution}
\end{figure}

In Fig.\,\ref{fig:ep_distribution}, we show
the posterior distributions of the $\epsilon$ for
\begin{align}
f_Q=(5,3,0) \ , \quad 
f_{\ubar} = (7,3,1)\ , \quad
f_{\dbar} = (6,5,5)\ ,
\label{eq:quark}
\end{align}
for the quark sector and 
\begin{align}
f_{L} = (5,4,4)\ , \quad
f_{\ebar} = (5,2,0)\ , \quad (\mbox{Lepton-1})\ ,
\label{eq:lepton1}
\end{align}
and 
\begin{align}
f_{L} = (2,0,-1)\ , \quad
f_{\ebar} = (10,7,5)\ , 
\quad (\mbox{Lepton-2})\ .
\label{eq:lepton2}
\end{align}
The figure shows that the $\epsilon$ distribution for the quark sector does not overlap much with that of the charge Lepton-1, while it overlaps well with that of the charge Lepton-2. 
This explains why, 
despite the charge Lepton-1 exhibiting a Bayes factor comparable to that of the charge Lepton-2 in separate analyses, 
the combined analysis results in the charge Quark+Lepton-2 having a higher Bayes factor than the charge Quark+Lepton-1 (see Tabs.\,\ref{tab:qres03}, \ref{tab:wres03} and \ref{tab:nongut_dim5_res03}).

\begin{table}[t!]
\caption{
Ten FN charge assignments with large Bayes factors for combined case with the seesaw mechanism.
These FN charges are arranged from the first generation to the third generation from left to right.
The range of $\epsilon$ is the 95\% CI of its posterior distribution.
}
 \begin{center}
  \begin{tabular}{|c|c|c|c|c|c|c|} \hline
    $f_Q$ & $f_{\ubar}$ & $f_{\dbar}$ & $f_L$ & $ f_{\ebar} $ & $\mathrm{log}_{10} (\calZ_{\mathrm{C}}/\calZ_{0,\mathrm{C}})$ & $\epsilon$ \\ \hline
    $5,3,0$ & $7,3,1$ & $6,5,5$ & $10,-9,-9$ & $1,3,5$ & $150.55\pm 0.03$ & $0.301 \to 0.322$\\
    $5,3,0$ & $8,4,1$ & $6,6,6$ & $10,-9,-9$ & $2,2,4$ & $150.49\pm 0.02$ & $0.336 \to 0.349$\\
    $5,3,0$ & $7,3,1$ & $6,5,5$ & $10,9,-9$ & $3,1,-5$ & $150.48\pm 0.02$ & $0.301 \to 0.321$\\
    $5,3,0$ & $7,3,1$ & $6,5,5$ & $3,-2,-2$ & $9,9,6$ & $150.42\pm 0.05$ & $0.307 \to 0.328$\\
    $5,3,0$ & $8,4,1$ & $6,6,6$ & $3,-2,-2$ & $9,9,7$ & $150.41\pm 0.02$ & $0.335 \to 0.349$\\
    $5,3,0$ & $7,3,1$ & $6,5,5$ & $5,5,4$ & $6,2,0$ & $150.39\pm 0.03$ & $0.303 \to 0.323$\\
    $5,3,0$ & $7,3,1$ & $6,5,5$ & $3,-2,-2$ & $9,8,6$ & $150.39\pm 0.03$ & $0.303 \to 0.324$\\
    $5,3,0$ & $8,4,1$ & $6,6,6$ & $6,5,5$ & $6,2,0$ & $150.26\pm 0.02$ & $0.335 \to 0.349$\\
    $5,3,0$ & $7,3,1$ & $7,5,5$ & $5,5,4$ & $6,2,0$ & $150.09\pm 0.04$ & $0.309 \to 0.326$\\
    $5,3,0$ & $7,3,1$ & $5,5,5$ & $5,5,4$ & $6,2,0$ & $150.08\pm 0.03$ & $0.300 \to 0.319$\\
 \hline
  \end{tabular}
  \label{tab:nongut_seesaw_res03}
 \end{center} 
\end{table}

\begin{table}[t!]
\caption{
Ten FN charge assignments with large Bayes factors for combined case with the dimension-five operators.
These FN charges are arranged from the first generation to the third generation from left to right.
The range of $\epsilon$ is the 95\% CI of its posterior distribution.
}
 \begin{center}
  \begin{tabular}{|c|c|c|c|c|c|c|} \hline
    $f_Q$ & $f_{\ubar}$ & $f_{\dbar}$ & $f_L$ & $ f_{\ebar} $ & $\mathrm{log}_{10} (\calZ_{\mathrm{C}}/\calZ_{0,\mathrm{C}})$ & $\epsilon$ \\ \hline
    $5,3,0$ & $8,4,1$ & $6,6,6$ & $2,0,-1$ & $10,8,5$ & $151.57\pm 0.02$ & $0.335 \to 0.349$\\
    $5,3,0$ & $7,4,1$ & $7,5,5$ & $2,0,-1$ & $10,7,5$ & $151.51\pm 0.02$ & $0.317 \to 0.338$\\
    $5,3,0$ & $7,3,1$ & $6,5,5$ & $5,5,4$ & $6,2,0$ & $151.50\pm 0.02$ & $0.303 \to 0.323$\\
    $5,3,0$ & $7,3,1$ & $7,5,5$ & $2,0,-1$ & $10,7,5$ & $151.49\pm 0.02$ & $0.312 \to 0.333$\\
    $5,3,0$ & $7,3,1$ & $5,6,5$ & $2,0,-1$ & $10,7,5$ & $151.49\pm 0.02$ & $0.308 \to 0.329$\\
    $5,3,0$ & $8,4,1$ & $6,6,6$ & $4,-1,-1$ & $8,8,6$ & $151.46\pm 0.02$ & $0.335 \to 0.349$\\
    $5,3,0$ & $8,4,1$ & $6,6,6$ & $5,5,4$ & $7,3,0$ & $151.43\pm 0.05$ & $0.335 \to 0.349$\\ 
    $5,3,0$ & $7,3,1$ & $6,5,5$ & $5,5,4$ & $6,2,0$ & $151.42\pm 0.03$ & $0.302 \to 0.323$\\
    $5,3,0$ & $8,4,1$ & $7,6,5$ & $5,5,4$ & $7,3,0$ & $151.24\pm 0.05$ & $0.333 \to 0.349$\\
    $5,3,0$ & $7,3,1$ & $5,5,5$ & $5,5,4$ & $6,2,0$ & $151.22\pm 0.02$ & $0.299 \to 0.319$\\
\hline
  \end{tabular}
  \label{tab:nongut_dim5_res03}
 \end{center} 
\end{table}

\subsubsection*{
Indication to SU(5) GUT consistent charge assignment}
\label{sec:gut_result}
\begin{figure}[t]
	\centering
	{\includegraphics[width=0.75\textwidth]{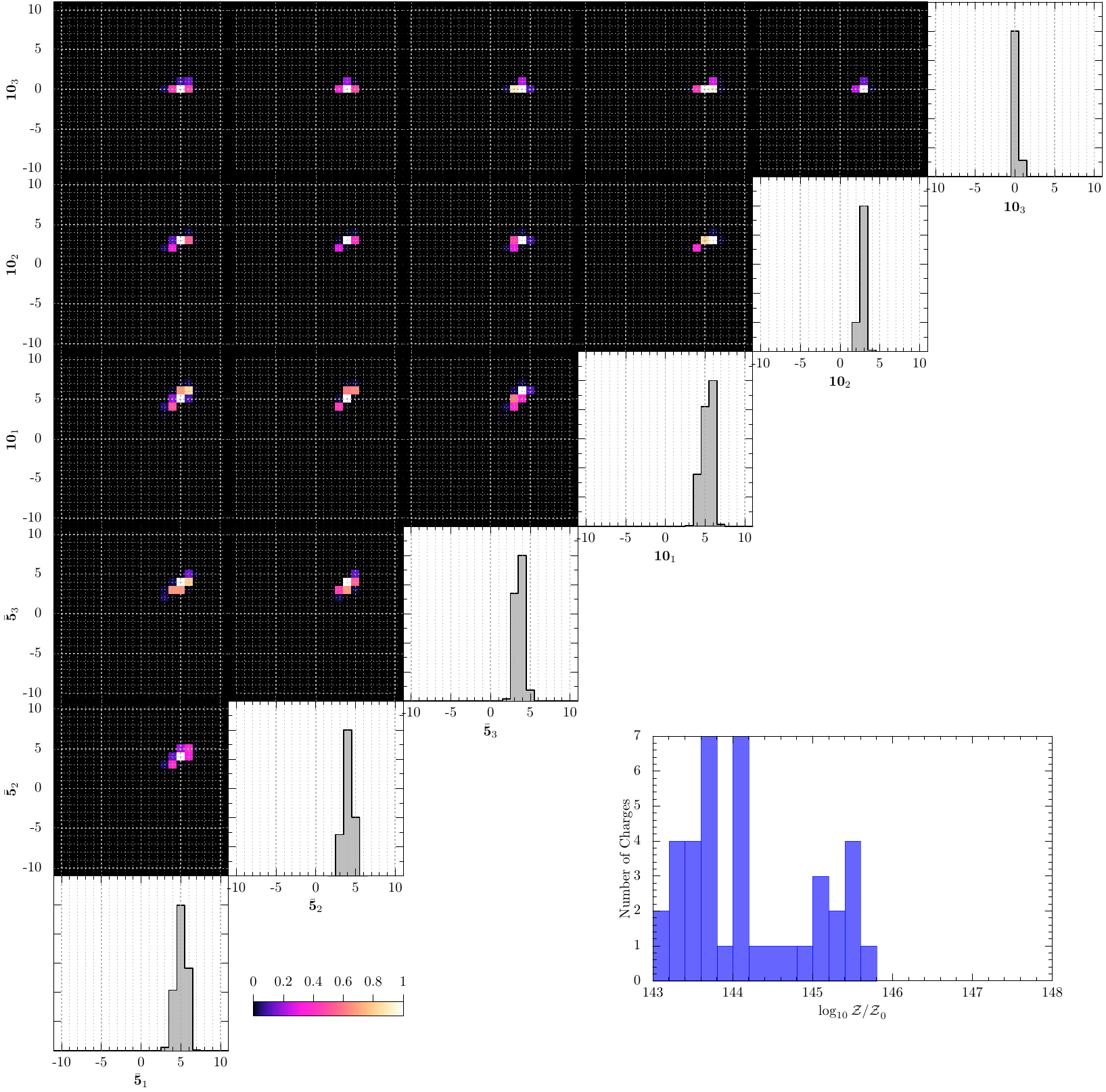}}
 \caption{The upper-left panel shows a two-dimensional histogram of GUT-like charges in the seesaw mechanism.}
 \label{fig:gut_seesaw_corner}
\end{figure}

\begin{figure}[t]
	\centering
	{\includegraphics[width=0.75\textwidth]{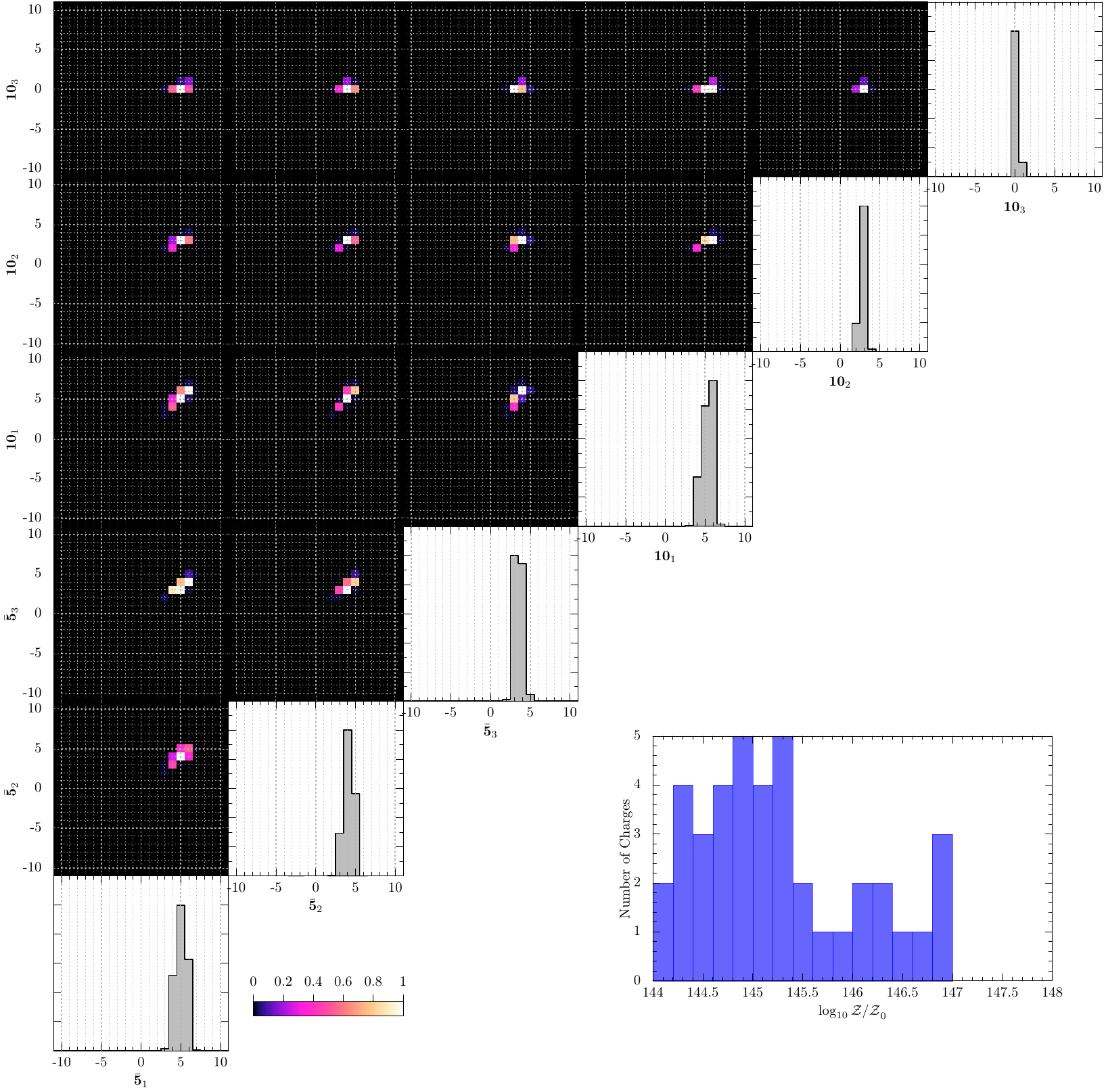}}
 \caption{The upper-left panel shows a two-dimensional histogram of GUT-like charges in the dimension-five operator case. }
 \label{fig:gut_dim5_corner}
\end{figure}

Let us consider the FN charge assignment which is consistent with the SU(5) GUT, 
that is,
\begin{gather}
f_{\overline{\mathbf{5}}}=f_{L}=f_{\dbar}\ , \quad
f_{\mathbf{10}}=f_{Q}=f_{\ubar}=f_{\ebar}\ .
\end{gather}
Tables.\,\ref{tab:gut_seesaw_res03} and \ref{tab:gut_dim5_res03}  
lists ten examples of GUT-like FN charges for each case: the seesaw mechanism and the dimension-five operator. 
The meaning of 
$\calZ_\GUT$, $\calZ_{0,\GUT}$ and $\epsilon$'s range are the same as before.
Fig.\,\ref{fig:gut_seesaw_corner} and \ref{fig:gut_dim5_corner} show two-dimensional histograms of the GUT-like FN charges 
for the seesaw mechanism and the dimension-five operator cases.

As a caveat of our results, 
we have not imposed the unification of the Yukawa coupling constants between the 
down-type quarks and the charged leptons.
As is well known, 
when the Yukawa couplings of the SM are extrapolated to the GUT scale, 
the unified Yukawa couplings for the charged leptons and down-type quarks predicted from the tree-level Yukawa couplings of SU(5) GUT do not match. 
These causes the lower Bayes factors in 
Tabs.\,\ref{tab:gut_seesaw_res03} and \ref{tab:gut_dim5_res03} lower than those in Tabs.\,\ref{tab:nongut_seesaw_res03} and \ref{tab:nongut_dim5_res03},
respectively.
Considering this, 
our results should be
taken as a demonstration, 
which roughly shows
the consistency of the FN mechanism with GUT. 
To explore the FN charge candidates for GUT,
Bayes factor analysis should be performed for each GUT model that can reproduce the SM Yukawa couplings (see e.g., Refs.\,\cite{Ibe:2019ifm,Ibe:2022ock,Ibe:2024cbt}).

\begin{table}[t!]
\caption{
Ten FN charge assignments with large Bayes factors for the seesaw mechanism.
These FN charges are arranged from the first generation to the third generation from left to right.
The range of $\epsilon$ is the 95\% CI of its posterior distribution.
}
 \begin{center}
  \begin{tabular}{|c|c|c|c|} \hline
    $f_{\overline{\mathbf{5}} \supset \dbar, L}$ & $f_{\mathbf{10} \supset Q, \ubar, \ebar}$ & $\mathrm{log}_{10} (\calZ_{\GUT}/\calZ_{0,\GUT})$ & $\epsilon$ \\ \hline
    $5,4,3$ & $5,3,0$ & $145.83\pm 0.06$ & $0.257 \to 0.274$\\
    $4,3,3$ & $4,2,0$ & $145.73\pm 0.04$ & $0.180 \to 0.199$\\
    $6,5,4$ & $6,3,0$ & $145.57\pm 0.07$ & $0.316 \to 0.340$\\
    $5,4,4$ & $5,3,0$ & $145.57\pm 0.05$ & $0.276 \to 0.293$\\
    $5,4,4$ & $6,3,0$ & $145.46\pm 0.05$ & $0.297 \to 0.317$\\
    $5,5,4$ & $6,3,0$ & $145.44\pm 0.05$ & $0.307 \to 0.329$\\
    $6,4,4$ & $6,3,1$ & $145.33\pm 0.06$ & $0.321 \to 0.343$\\
    $4,4,3$ & $5,3,0$ & $145.20\pm 0.06$ & $0.246 \to 0.268$\\
    $6,5,5$ & $6,3,0$ & $145.15\pm 0.16$ & $0.336 \to 0.349$\\
    $6,4,4$ & $6,3,0$ & $145.13\pm 0.06$ & $0.309 \to 0.327$\\ \hline 
  \end{tabular}
  \label{tab:gut_seesaw_res03}
 \end{center} 
\end{table}

\begin{table}[t!]
\caption{
Ten FN charge assignments with large Bayes factors for dimension-five operators.
These FN charges are arranged from the first generation to the third generation from left to right.
The range of $\epsilon$ is the 95\% CI of its posterior distribution.
}
 \begin{center}
  \begin{tabular}{|c|c|c|c|} \hline
    $f_{\overline{\mathbf{5}} \supset \dbar, L}$ & $f_{\mathbf{10} \supset Q, \ubar, \ebar}$ & $\mathrm{log}_{10} (\calZ_{\GUT}/\calZ_{0,\GUT})$ & $\epsilon$ \\ \hline
    $5,4,3$ & $5,3,0$ & $147.16\pm 0.06$ & $0.252 \to 0.274$\\
    $4,3,3$ & $4,2,0$ & $146.94\pm 0.04$ & $0.179 \to 0.197$\\
    $6,5,4$ & $6,3,0$ & $146.90\pm 0.06$ & $0.317 \to 0.338$\\
    $5,5,4$ & $6,3,0$ & $146.84\pm 0.03$ & $0.305 \to 0.329$\\
    $4,4,3$ & $5,3,0$ & $146.69\pm 0.06$ & $0.244 \to 0.266$\\
    $6,4,4$ & $6,3,1$ & $146.56\pm 0.07$ & $0.321 \to 0.341$\\
    $6,4,4$ & $6,3,0$ & $146.39\pm 0.06$ & $0.310 \to 0.326$\\
    $5,4,4$ & $5,3,0$ & $146.31\pm 0.06$ & $0.279 \to 0.296$\\
    $5,4,4$ & $6,3,0$ & $146.18\pm 0.04$ & $0.296 \to 0.317$\\
    $6,5,5$ & $6,3,0$ & $146.09\pm 0.13$ & $0.331 \to 0.349$\\ \hline 
  \end{tabular}
  \label{tab:gut_dim5_res03}
 \end{center} 
\end{table}

\section{Predictions of
Lightest Neutrino Mass and Neutrinoless Double Beta Decay}
\label{sec:mbb}

So far, we have used dimensionless observables to discuss the goodness of the FN charge assignments.
In this section, we make predictions about the lightest neutrino mass scale and neutrinoless double beta decay rate based on the posterior distributions obtained in previous section.
\subsection{Posterior distribution of 
\texorpdfstring{$m_1$}{} 
and 
\texorpdfstring{$m_{ee}$}{} 
}
\label{sec:mbb_setup}
To obtain the posterior distribution of $m_{1}$
and $m_{ee}$
let us focus on the 
posterior distribution which is projected onto the $\hat{y}_{1}^2 := m_{1}^2M_R^2/v_\mathrm{EW}^4 $ and $\Delta \hat{y}_{31}^2:=\Delta m_{31}^{2}M_R^2/v_\mathrm{EW}^4$,
i.e., 
\begin{align}
\int d \hat{y}_{1}^{2}\,
d \Delta \hat{y}_{31}^{2}\, 
\, \mathrm{P}(\hat{y}_{1}^{2}, \Delta \hat{y}_{31}^{2}) \ .
\end{align}
Here, we assume the seesaw mechanism
(see Eq.\,\eqref{eq:seesaw_LY}).
When the neutrino masses originate from the dimension-five operators in Eq.\,\eqref{eq:wop}, $M_R$ is replaced by $\Lambda_W$.
The posterior distribution of the lightest neutrino mass squared, $m_1^2$, is obtained by fixing the right-handed neutrino mass scale, 
i.e., $M_R$ (or $\Lambda_W$).
As the prior distribution
of $M_R$, 
we assume it by
\begin{align}
    \pi(M_R)\propto M_R^p\ ,
\end{align}
where we vary the exponent $p$ in $p=(-2,2)$ to estimate the uncertainty 
of the prediction due to the choice of the prior distribution.
By noting that $\Delta m_{31}^{2}$ is precisely measured, we obtain the 
posterior distribution of $m_1^2$ 
from, 
\begin{align}
&\int d \hat{y}_{1}^{2}\,
d \Delta \hat{y}_{31}^{2}\,  dM_R\, 
\mathrm{P}(\hat{y}_{1}^{2}, \Delta \hat{y}_{31}^{2})\, \pi(M_R)\, \delta \left(\Delta m_{31}^{2\mathrm{(obs)}} - \frac{\Delta \hat{y}_{31}^{2} v_\mathrm{EW}^{4}}{M_{R}^{2}} \right) \  \nonumber \\
&= 
\int d m_{1}^{2}\, d \Delta \hat{y}_{31}^{2}\,
d M_{R}\,  
\frac{M_{R}^{2}}{v_\mathrm{EW}^{4}}\,
\mathrm{P}( m_{1}^{2}M_R^2/v_\mathrm{EW}^4, \Delta \hat{y}_{31}^{2})\, 
\pi(M_{R})\, 
\delta \left(\Delta m_{31}^{2\mathrm{(obs)}} - \frac{\Delta \hat{y}_{31}^{2} v_\mathrm{EW}^{4}}{M_{R}^{2}} \right) \ . 
\end{align}
By integrating over $M_R>0$,
the posterior distribution of $m_{1}^2$ is given by,
\begin{align}
    \int dm_{1}^2\, \mathrm{P}(m_1^2)\ ,
\end{align}
where
\begin{align}
\mathrm{P}(m_1^2) 
&\propto 
\int d \Delta \hat{y}_{31}^{2}\,
(\Delta \hat{y}_{31}^{2})^{(3+p)/2}\,
\mathrm{P}( 
\Delta \hat{y}_{31}^2m_{1}^{2} /\Delta m_{31}^{2\mathrm{(obs)}}, \Delta \hat{y}_{31}^{2}) \ .
\label{eq:m1_distribution}
\end{align}
Here, we have ignored the normalization of $\mathrm{P}(m_1^2)$.

The posterior distribution of $(m_1,m_{ee})$ can also be obtained by 
using 
\begin{align}
\int d m_{1}\, dm_{ee}\,
 \mathrm{P}(m_{1},m_{ee})
\propto \int d m_{1}\, dm_{ee}\,
d\eta_{1}\,
d\eta_{2}\, 
\delta(m_{ee}-m_{ee}^{(\mathrm{th})}(m_1,\eta_1,\eta_2))
\, \mathrm{P}(m_{1})\ ,
\end{align}
where $\eta_{1,2}\in[0,2\pi)$ are the  Majorana phases.
The theoretical expression of $m_{ee}^{(\mathrm{th})}$ for given 
$(m_1,\eta_1,\eta_2)$ is given in Ref.\,\cite{ParticleDataGroup:2022pth}.
Notice that the prior distributions 
of the Majorana phases are flat in $[0,2\pi)$.
Since all the observations so far are independent of the Majorana phases, the posterior distributions of them are also flat.

\subsection{Result}
\label{sec:mbb_result}
\begin{figure}[t!]
    \centering
    \begin{subfigure}{0.49\textwidth}
        \centering
        \includegraphics[width=\textwidth]{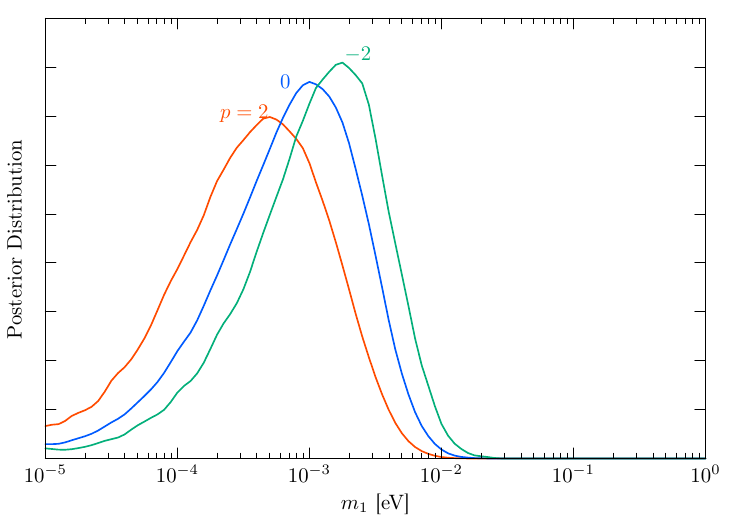}
        \caption{For the seesaw mechanism.}
        \label{fig:numass_power_seesaw}
    \end{subfigure}
    \hfill
    \begin{subfigure}{0.49\textwidth}
        \centering
        \includegraphics[width=\textwidth]{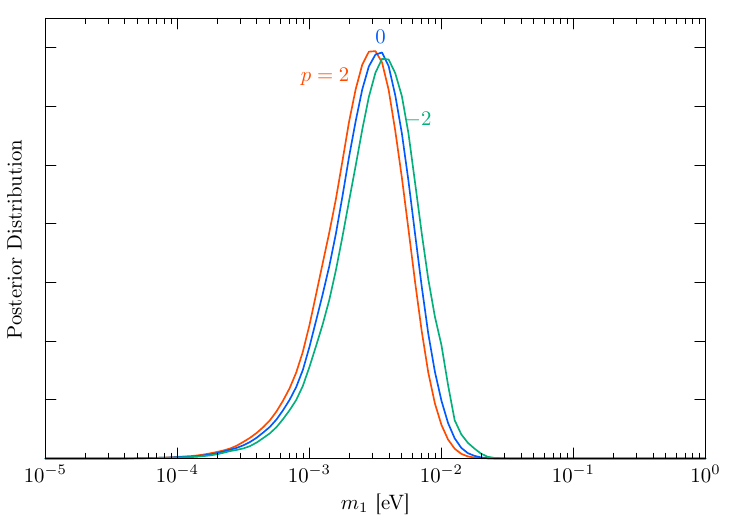}
        \caption{For the $ \mathrm{dimension\mathchar`-five}$ operator.}
        \label{fig:numass_power_wop}
    \end{subfigure}

\caption{
Posterior distribution of the lightest neutrino mass.
Index $p$ denotes the exponent of the prior distribution of $M_R$ or $\Lambda_W$.}
  \label{fig:neutrinomass_posterior}
\end{figure}

In Fig.\,\ref{fig:neutrinomass_posterior}, 
we show the probability density function (PDF) of the lightest neutrino mass, $\mathrm{P}(m_1)$ obtained by integrating the expression in Eq.\,\eqref{eq:m1_distribution}
for the seesaw mechanism and the dimension-five operator case.
We take the following FN breaking parameter $\epsilon$ and FN charge assignments,
\begin{align}
\label{eq:BP_seesaw}
&\epsilon = 0.28\ , \quad
f_{L} = (5,4,4)\ , \quad
f_{\ebar} =(5,2,0)\ , \quad \mathrm{(for\,\, the\,\, seesaw\,\, mechanism)}\ , \\
\label{eq:BP_dim5}
&\epsilon = 0.26 \ , \quad
f_{L} = (5,4,4)\ , \quad
f_{\ebar} =(5,2,0)\ , \quad \mathrm{(for\,\, the\,\, dimension\mathchar`-five\,\, operator)}\ ,
\end{align}
in the figure.
Here, to examine the differences in predictions between the seesaw mechanism and the dimension-five operator, we selected examples of the FN charges that are common to both Tabs.\,\ref{tab:sres03} and \ref{tab:wres03}.
The figure shows that the distribution 
peaks at around $m_1 \sim \sqrt{\Delta m^{2}_{21}}$.
This peak arises because realizing 
$m_1 \gg \sqrt{\Delta m^2_{21}}$ requires tuning, while the parameter space volume is small in the region where $m_1 \ll \sqrt{\Delta m^2_{21}}$. 
Furthermore, it is understood that this tendency does not strongly depend on the power-law of the right-handed neutrino mass distribution.
Notice that the predicted range of $m_1$ is below the current limit from the cosmological observations (see e.g., Ref.\,\cite{Loureiro:2018pdz}).

\begin{figure}[t!]
    \centering
    \begin{subfigure}[b]{0.49\textwidth}
        \centering
        \includegraphics[width=\textwidth]{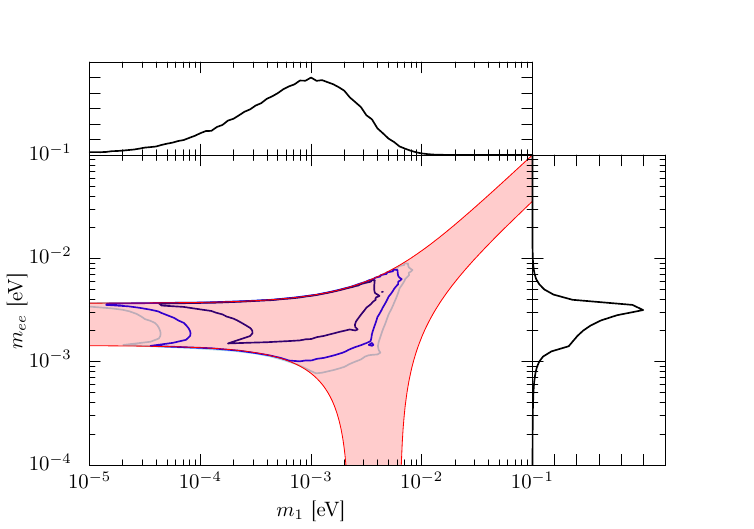}
        \caption{For the seesaw mechanism.}
    \end{subfigure}
    \hfill
    \begin{subfigure}[b]{0.49\textwidth}
        \centering
        \includegraphics[width=\textwidth]{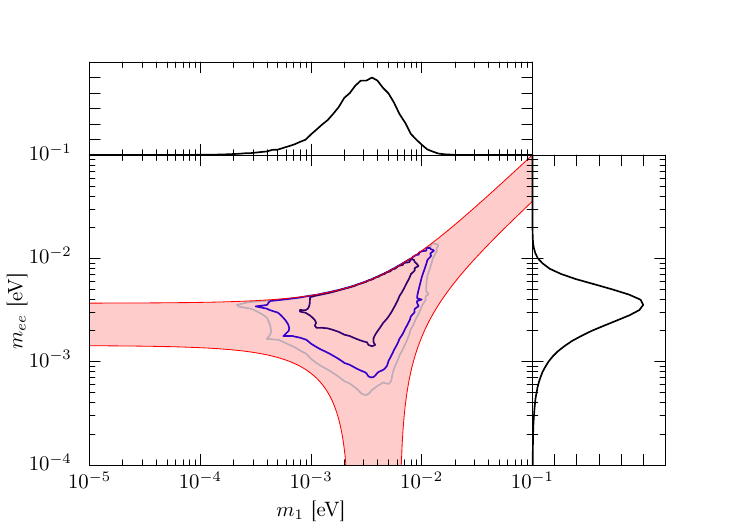}
        \caption{For the $ \mathrm{dimension\mathchar`-five}$ operator.}
    \end{subfigure}
    
\caption{
Posterior distribution of $m_{ee}$. 
The exponent of the prior distribution of $M_R$ or $\Lambda_W$ is set to 0.}
\label{fig:mee_posterior}
\end{figure}

In Fig.\,\ref{fig:mee_posterior}, 
we show the contour plots of $\mathrm{P}(m_1,m_{ee})$.
The results indicate that there is also a region where $m_{ee}\gtrsim 10^{-2}$\,eV. 
Such a region could be tested by next-generation experiments such as the LEGEND-1000 experiment\,\cite{LEGEND:2021bnm}. 
On the other hand, 
in most of the parameter space, 
$ m_{ee}\gtrsim 10^{-3}$\,eV. 
Exploring these regions requires further scaled-up experiments.

\begin{figure}[t!]
	\centering	{\includegraphics[width=0.55\textwidth]{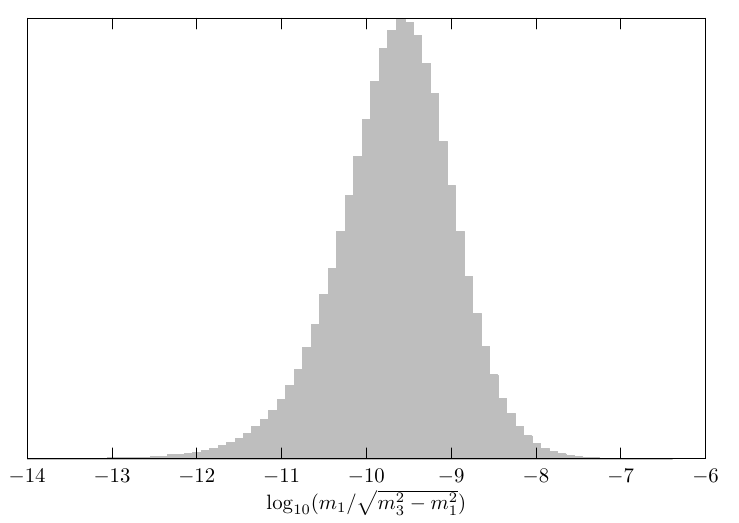}}
 \caption{The prior distribution of the masses of the lightest and heaviest neutrinos in the seesaw mechanism for the case where $\epsilon = 0.28, f_L = (8,0,0), f_{\ebar} = (10,6,-4)$.}
 \label{fig:lightest_nu_mass}
\end{figure}

Finally, 
it should be noted that even in the FN mechanism, 
specific charge assignments can induce a hierarchical mass structure in the neutrino masses, 
making the lightest neutrino mass significantly smaller than the other two.
As the charge distributions in Figs.\,\ref{fig:seesaw_corner} and \ref{fig:dim5_corner}
show, 
there are many cases where only $|f_{L,1}|$ takes a large value, 
while $|f_{L,2}|$ and $|f_{L,3}|$ are not large.
For such charge assignments,  
the rank of the effective neutrino mass matrix is reduced to two due to the   
large $|f_{L,1}|$.
As a result, one of the neutrino being nearly massless
for such charge assignments.
Fig.\,\ref{fig:lightest_nu_mass} shows the prior distribution of the relation of masses of the lightest and heaviest neutrinos in the seesaw mechanism for the case where $\epsilon = 0.28, f_L = (8,0,0), f_{\ebar} = (10,6,-4)$.
From this figure, 
it can be seen that the lightest neutrino has a significantly smaller mass compared to the other generations of neutrinos.

\section{Implication of Nucleon Decay}
\label{sec:neuleon_decay}

The FN mechanism provides a consistent framework for explaining 
flavor structures of 
the SM.
Furthermore, this mechanism offers a rich avenue for exploring new physics beyond the SM, 
particularly in the context of flavor-changing processes.
For example, in the supersymmetric Standard Model,
not only the flavor structures of the fermion but also those of the scalar fermions are affected by the FN mechanism.
In such cases, 
several flavor changing effects are affected\,\cite{Murayama:1994tc,Harnik:2004yp,Gabbiani:1996hi,Altmannshofer:2013lfa,Nagata:2013sba}.

Among the various new physics implications influenced by the FN mechanism, we focus here specifically on nucleon decay. 
Flavor structures at a high energy affect the nucleon's decay rate.
Particularly,
branching fraction is an important probe of origins of flavor structures.
In fact, 
it is known that the flavor structure has a significant impact on the nucleon lifetime and decay modes arising from dimension-five operators in supersymmetric models\,\cite{Dine:2013nga,Nagata:2013sba}.
It has also been pointed out that the flavor structure of leptons has a significant impact on nucleon decay in models known as fake GUTs\,\cite{Ibe:2019ifm,Ibe:2022ock,Ibe:2024cbt}.  
We have explored the FN charge in SM without supersymmetry. Therefore, 
we now focus on nucleon decay in the non-supersymmetric case.
We will address the selection of FN charges and their impact on nucleon decay in future work within the supersymmetric SM\,\cite{CISW2024}.

In this paper,
we discuss nucleon decay within a broad framework, 
without being restricted to GUTs.
For this purpose,
an analysis based on effective operators is particularly useful.
Assuming the SM is valid up to some cutoff scale, 
we consider 
the following dimension-six operators which cause baryon number-violating nucleon decay%
\footnote{We adopt the notation used in Ref.\,\cite{Nagata:2013sba}.}
\cite{Weinberg:1979sa,Wilczek:1979hc,Abbott:1980zj}:
\begin{align}
\label{eq:SM_dim6_1}
\calO_{ijkl}^{(1)} 
&= 
\epsilon^{\alpha \beta} \epsilon_{abc} 
( \bar{u}^{\dagger a}_i\, \bar{d}^{\dagger b}_j ) 
( Q_{k \alpha}^c\, L_{l \beta} )\ , \\
\label{eq:SM_dim6_2}
\calO_{ijkl}^{(2)} 
&= 
\epsilon^{\alpha \beta} \epsilon_{abc} 
( Q_{i \alpha}^a\, Q_{j \beta}^b )
(\bar{u}^{\dagger c}_k\, \bar{e}^{\dagger}_l)\ , \\ 
\label{eq:SM_dim6_3}
\calO_{ijkl}^{(3)} 
&= 
\epsilon^{\alpha \beta} \epsilon^{\gamma \delta} \epsilon_{abc} 
( Q_{i \alpha}^a\, Q_{j \gamma}^b ) ( Q_{k \delta}^c\, L_{l \beta} )\ , \\
\label{eq:SM_dim6_4}
\calO_{ijkl}^{(4)} 
&= 
\epsilon_{abc}  
(\bar{u}^{\dagger a}_i\, \bar{d}^{\dagger b}_j)
(\bar{u}^{\dagger c}_k\, \bar{e}^{\dagger}_l)\ ,
\end{align}
where $a,b,c$ denote the $\SU(3)_c$ indices
and $\alpha,\beta$ represent the $\SU(2)_L$ indices.
Below the $\UFN$ symmetry breaking scale,
we can expect following patterns of $\epsilon$ dependence of these operators,
\begin{align}
C^{ijkl}_{(1)} 
&= \frac{A^{ijkl}_{(1)}}{M_{(1)}^2} \times \epsilon^{| - f_{\ubar,i} - f_{\dbar,j} + f_{Q,k} + f_{L,l} |}\ , \\ 
C^{ijkl}_{(2)}  
&= \frac{A^{ijkl}_{(2)}}{M_{(2)}^2} \times \epsilon^{| f_{Q,i} + f_{Q,j} - f_{\ubar,k} -f_{\ebar,l} |}\ , \\ 
C^{ijkl}_{(3)}  
&= \frac{A^{ijkl}_{(3)}}{M_{(3)}^2} \times \epsilon^{| f_{Q,i} + f_{Q,j} + f_{Q,k} + f_{L,l} |}\ , \\
C^{ijkl}_{(4)}  
&= \frac{A^{ijkl}_{(4)}}{M_{(4)}^2} \times \epsilon^{| - f_{\ubar,i} - f_{\dbar,j}
 -f_{\ubar,k} -f_{\ebar,l} |}\ .
\end{align}
Here,
$M_{(1\mathchar`-4)}$ are 
flavor independent high energy scales. 
$A^{ijkl}_{(1\mathchar`-4)}$ are $\order{1}$ complex coefficients.
In the following, we take $M_{(1\mathchar`-4)}$ as free parameters,
while we assume $A^{ijkl}_{(1\mathchar`-4)}$ 
follow distributions similar to those of $\kappa$'s (see Eq.\,\eqref{eq:distribution_1}).

\begin{table}[t!]
\caption{Four benchmark points of FN charge assignments used for nucleon decay calculations.
}
 \begin{center}
  \begin{tabular}{|c|c|c|c|c|c|c|} \hline
    & $f_Q$ & $f_{\ubar}$ & $f_{\dbar}$ & $f_L$ & $ f_{\ebar} $ & $\epsilon$ \\ \hline
    $\mathrm{BP0}$ & $0,0,0$ & $0,0,0$ & $0,0,0$ & $0,0,0$ & $0,0,0$ & $ - $ \\
    $\mathrm{BP1}$ & $4, 2, 0$ & $4, 2, 0$ & $4, 3, 3$ & $4, 3, 3$ & $4, 2, 0$ & $ 0.185 $ \\
    $\mathrm{BP2}$ & $5, 3, 0$ & $7, 3, 1$ & $6, 5, 5$ & $5, 5, 4$ & $6, 2, 0$ & $0.31$ \\
    $\mathrm{BP3}$ & $5, 3, 0$ & $7, 3, 1$ & $6, 5, 5$ & $-5, -5, -4$ & $-6, -2, 0$  & $0.31$ \\ \hline
  \end{tabular}
\label{tab:charge_nucleon_decay}
\end{center} 
\end{table}

In the following, 
we consider four benchmark points
with the seesaw mechanism,
BP0, BP1, BP2 and BP3 in Tab.\,\ref{tab:charge_nucleon_decay}.
The point BP0 corresponds to no FN charge.
The points BP1 and BP2 are taken from Tabs.\,\ref{tab:nongut_seesaw_res03} and \ref{tab:gut_seesaw_res03}, which have 
large Bayes factors.
The point BP3 is obtained by flipping the sign of the lepton charge in BP2 and has the same Bayes factor as BP2.

To predict nucleon decay rates, 
we follow the Monte Carlo procedure outlined below.
For a given set of the FN breaking parameters and FN charges, 
we first generate $\kappa$ values and $A^{(ijkl)}_{(1\mathchar`-4)}$ by sampling from the Gaussian distributions as in Eqs.\,\eqref{eq:distribution_1} and \eqref{eq:distribution_2}.
Subsequently, 
using the $\kappa$ values and the given $A^{(ijkl)}_{(1\mathchar`-4)}$ that reproduce the SM fermion parameters, 
we calculate the nucleon decay rates for each decay mode for given realizations.

For each realization of $\kappa$'s,
we first transform the random flavor basis of SM fermions to their mass basis where 
the up-type and electron-type Yukawa matrices are diagonal,
and 
the Yukawa matrices of down-type and neutrino mass dimension-five operators are is given by
the diagonal matrix multiplied by the CKM and PMNS matrices,
respectively.
For the matrix elements for nucleon decay,
we utilize the central values of them by the QCD lattice calculation
in Refs.\,\cite{Yoo:2021gql,Aoki:2017puj}.
The uncertainty in the matrix elements is sufficiently smaller than the width of the prediction range shown in the following figures.

\begin{figure}[t!]
\centering

\raisebox{50pt}{\subcaptionbox{$\calO^{(1)}$
\label{fig:BP1}}
{\includegraphics[width=0.49\textwidth]{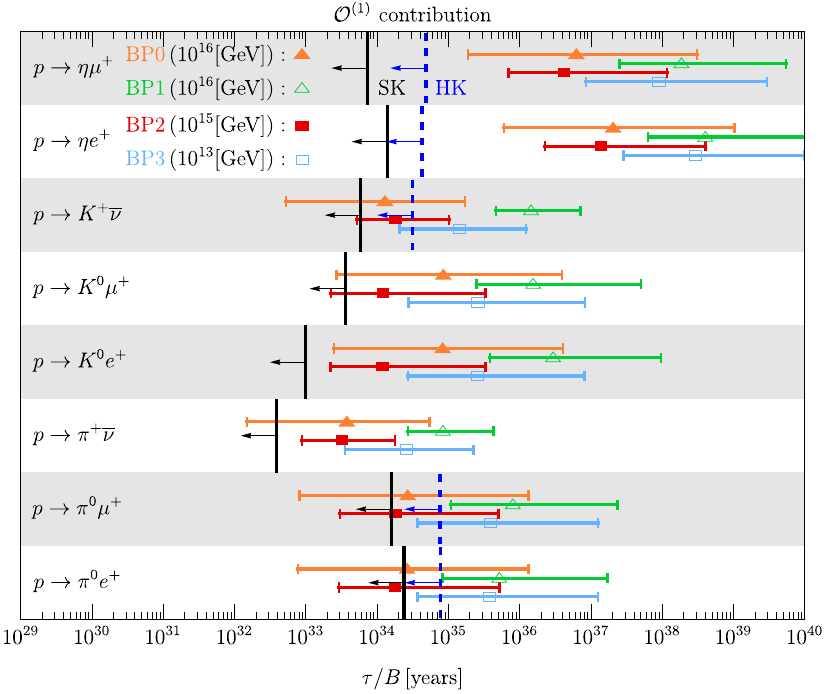}}}
\raisebox{50pt}{
\subcaptionbox{$\calO^{(2)}$
\label{fig:BP2}}
{\includegraphics[width=0.49\textwidth]{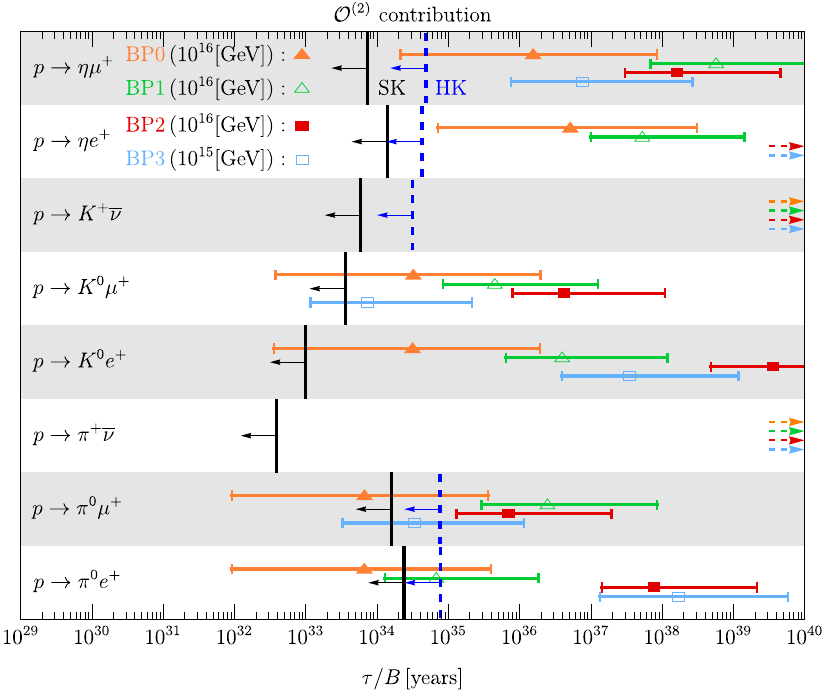}}}
\subcaptionbox{$\calO^{(3)}$
\label{fig:BP3}}
{\includegraphics[width=0.49\textwidth]{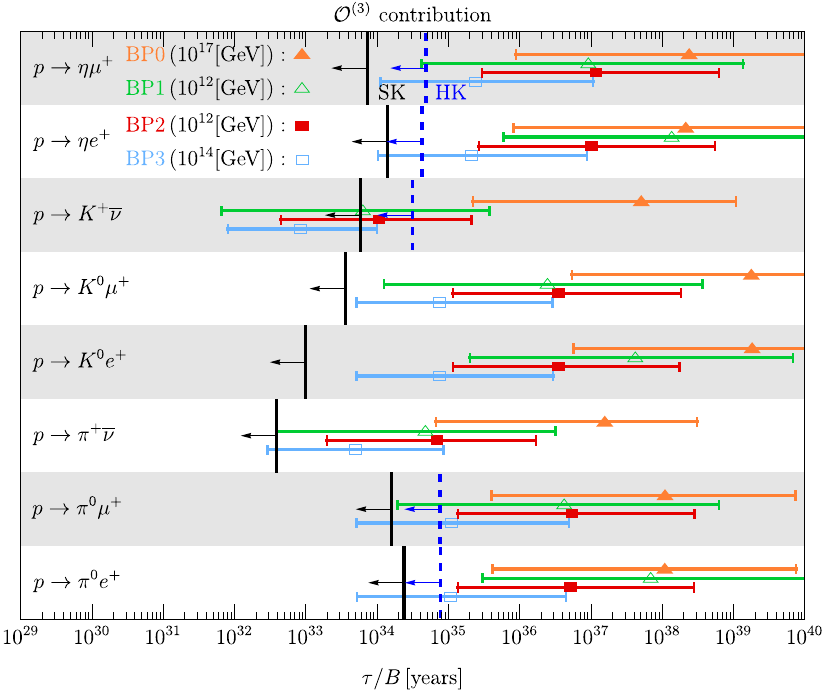}}
\subcaptionbox{$\calO^{(4)}$
\label{fig:BP4}}
{\includegraphics[width=0.49\textwidth]{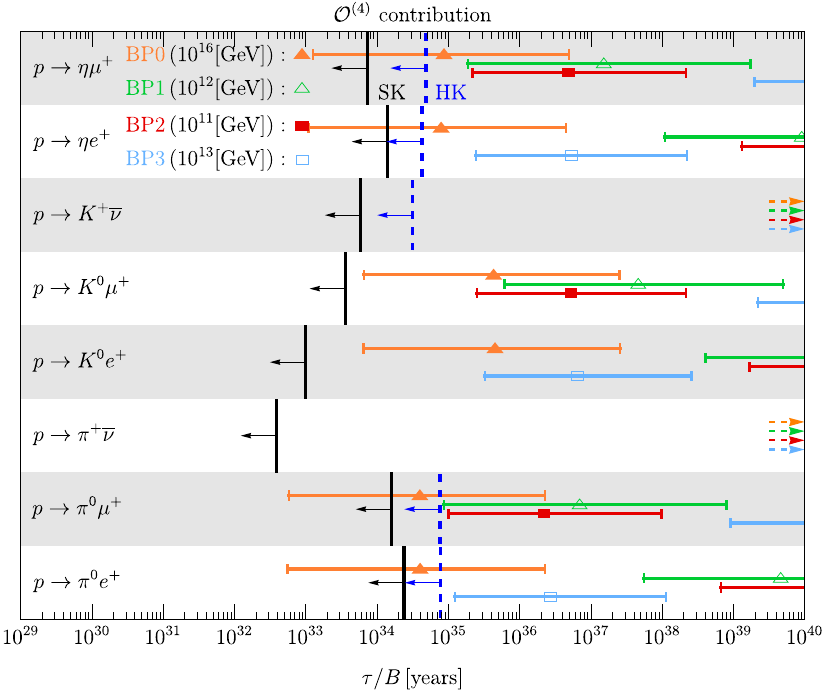}}
\caption{The nucleon lifetimes for various decay modes in four benchmark points. 
Predictions in BP0, BP1, BP2 and BP3 are shown by orange bands, 
green bands, 
red bands,
and light blue bands,
respectively.
These horizontal bars represent $2 \sigma$ ranges of the lifetime distributions, 
with median values indicated by markers of each color.
We set the values of the cutoff scales $M_{(1\mathchar`-4)}$ to those indicated in the figure for each benchmark point.
The colored dashed rightward arrows indicate that the lifetime is longer than $10^{40}$ years.
Current constraints from SK are shown with leftward black arrows\,\cite{Super-Kamiokande:2020wjk,Super-Kamiokande:2013rwg,Super-Kamiokande:2005lev,Super-Kamiokande:2022egr,Super-Kamiokande:2014otb,Super-Kamiokande:2024qbv}, 
and future projections for HK are indicated with leftward blue arrows\,\cite{Hyper-Kamiokande:2018ofw}.
}
\label{fig:dim6_nucleon_decay}
\end{figure}

In Fig.\,\ref{fig:dim6_nucleon_decay}, 
we present the percentile ranges for the predicted nucleon lifetimes across various decay modes for the FN charge assignments in Tab.\,\ref{tab:charge_nucleon_decay}.
Here, 
we set the values of the cutoff $M_{(1\mathchar`-4)}$ to those indicated in the figure for each benchmark point.
For decay modes involving $\overline{\nu}$,
we sum over the contributions from all neutrino flavors.
Predictions in each benchmark point are shown by orange bands, 
green bands, 
red bands,
and light blue bands,
respectively.
These bands represent $2 \sigma$ ranges of the lifetime distributions arising from uncertainties in the $\order{1}$ coefficients $\kappa$'s and $A$'s.
The markers indicate median values.
The colored dashed rightward arrows indicate that the lifetime is longer than $10^{40}$ years.
Black and blue leftward arrows indicate current constraints from SK\,\cite{Super-Kamiokande:2020wjk,Super-Kamiokande:2013rwg,Super-Kamiokande:2005lev,Super-Kamiokande:2022egr,Super-Kamiokande:2014otb,Super-Kamiokande:2024qbv} and the future projections for Hyper-Kamiokande (HK)\,\cite{Hyper-Kamiokande:2018ofw}.

The figures reveal that the nucleon lifetimes from baryon number violating operators in Eqs.\,\eqref{eq:SM_dim6_1}-\eqref{eq:SM_dim6_4},
have strong dependence on the FN charges. 
The nucleon lifetime varies significantly across all benchmark points.
These results highlight the critical role that future experiments probing diverse nucleon decay modes can play in finding FN mechanism.

\begin{figure}[t!]
	\centering
	{\includegraphics[width=0.55\textwidth]{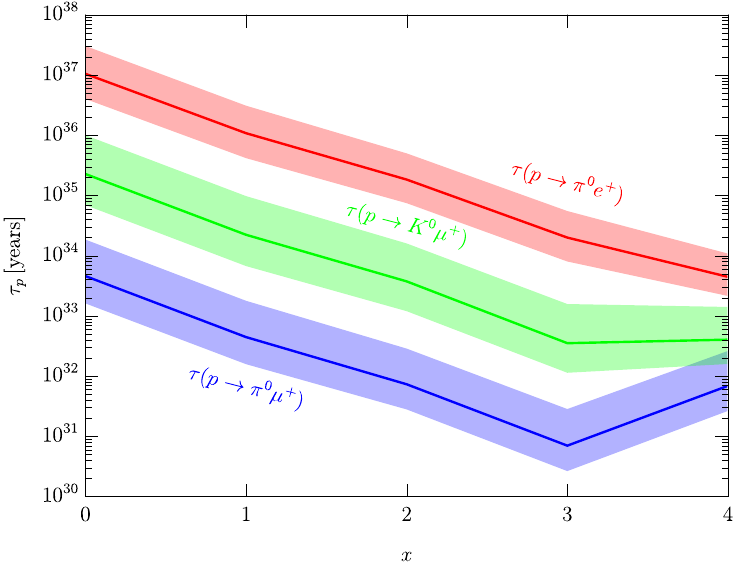}}
 \caption{Nucleon lifetimes for $f_L=(1+x,1+x,x)$ and $f_{\ebar}=(10-x, 6-x, 4-x)$.
 For $\epsilon$ and the FN charges of quarks, the same values as those in BP2 are used.}
 \label{fig:plifetime_x_change}
\end{figure}

As discussed, there is redundancy in the choice of FN charges that reproduce the physical parameters. 
In addition to the sign differences exemplified  
between BP2 and BP3, 
the charge shifts,
$f_L=(1+x,1+x,x)$ and $f_{\ebar}=(10-x, 6-x, 4-x)$, realize the same SM flavor structure for $x$ within the range of $0$ to $4$. 
However, 
such differences in charges do have an impact on nucleon decay.

In Fig.\,\ref{fig:plifetime_x_change}, 
we present the lifetime $\tau(p \to \pi^0 e^+)$,
$\tau(p \to \pi^0 \mu^+)$
and 
$\tau(p \to K^0 \mu^+)$,
involving contributions of all four decay operators in Eqs.\,\eqref{eq:SM_dim6_1}-\eqref{eq:SM_dim6_4}. 
These lifetimes are presented for the FN charges $f_L=(1+x,1+x,x)$ and $f_{\ebar}=(10-x, 6-x, 4-x)$. 
For $\epsilon$ and the FN charges of quarks, the same values as those in BP2 are used.
We set the value of common cutoff scale $M_{(1\mathchar`-4)} = 10^{15}$ GeV.
The lifetime bands depicted in the figure represent the $1 \sigma$ ranges.
From the figure, 
it is evident that varying $x$ significantly affects the lifetime.
Furthermore, 
the decay branching ratios are also greatly impacted.

\section{Conclusion}
\label{sec:conclusion}
In this paper, we discussed ``good" FN charge assignments up to $|f_{Q,\ubar,\dbar,L,\ebar}| \le 10$ using Bayesian inference. We began by analyzing the quark and lepton sectors separately. In the lepton sector specifically, we examined two approaches to neutrino mass generation: the seesaw mechanism and dimension-five effective operators.

We first conducted an analysis of the quark sector and identified approximately $\order{10^3}$ viable FN charge assignments. 
Among these, 
some had been previously studied in existing literature, 
while others included newly discovered, 
unconventional FN charges with negative values.

We performed a similar analysis for the lepton sector.
Our results showed that, 
in the lepton sector, 
charge assignments with $\order{10^3}$ candidates demonstrated a significantly large Bayes factor. 
Additionally, 
our analysis revealed that both mechanisms for neutrino mass generation provided comparable explanations for the flavor structure, 
highlighting no substantial differences between them.

For the good FN charge assignments of the lepton sector, 
we also discussed the lightest neutrino mass and neutrinoless double beta decay under the FN mechanism. 
When all FN charges of the lepton sector are non-negative, we found that the posterior distribution of the lightest neutrino mass peaks at around $m_1 \sim \sqrt{\Delta m_{21}^2}$ for both the seesaw mechanism and the dimension-five operator cases. 
This tendency does not strongly depend on the power-law of the right-handed neutrino mass scale $M_R$ (or $\Lambda_W$). 
Additionally, 
our analysis revealed that a parameter region exists where $m_{ee} \gtrsim 10^{-2}$\,eV for both methods of neutrino mass generation.

On the contrary, 
if negative FN charges are allowed in the seesaw mechanism, 
we found that the flavor structure of the lepton sector can still be reproduced with large $f_{L,1,2,3}$ values of different signs. 
In this case, 
one of the charged leptons decouples from the seesaw mechanism, 
resulting in an extremely small lightest neutrino mass. 
Neutrino masses will be further elucidated through future cosmological observations as well as neutrinoless double beta decay experiments. 
Our analysis revealed that information about neutrino masses is crucial for understanding the FN mechanism.

Additionally, 
to demonstrate that FN mechanism can simultaneously explain the flavor structures of both quarks and leptons, 
we computed the Bayes factor for the combined quark and lepton sectors.
As a result, 
we identified numerous FN charge assignments that effectively account for the flavor structure in the unified framework.

We analyzed nucleon decay induced by dimension-six effective operators using the FN charges identified in this study, which successfully explain the flavor structures of both quarks and leptons simultaneously.
Notably, 
even charge assignments that similarly explained the flavor structure led to significantly varied predictions for nucleon decay. 
These findings highlight the possibility of investigating flavor symmetries through nucleon decay observations.

\section*{Acknowledgements}
This work is supported by Grant-in-Aid for Scientific Research from the Ministry of Education, Culture, Sports, Science, and Technology (MEXT), Japan, 21H04471, 22K03615, 24K23938 (M.I.), 20H01895 and 20H05860  (S.S.) and by World Premier International Research Center Initiative (WPI), MEXT, Japan. 
This work is supported by JST SPRING, Grant Number JPMJSP2108 and ANRI fellowship (K.W.).

\appendix

\section{Physical Parameters}
\label{sec:msbar}

Here, 
we list the set of $\overline{\mathrm{MS}}$ parameters at $\mu_{R} = 10^{15}\, \mathrm{GeV}$ which we have used in our calculation. 
We have used the PDG average of the SM particle masses \cite{ParticleDataGroup:2022pth} and obtain relevant running parameters in 
the $\overline{\mathrm{MS}}$ following Ref.\,\cite{Buttazzo:2013uya}.
We estimate the Yukawa couplings of the light quarks, by using the QCD four-loop RGEs and three-loop decoupling effects from heavy quarks \cite{Chetyrkin:1997sg}.

\begin{table}[t!]
\caption{The SM parameters 
determined from the observations 
are given in $\overline{\mathrm{MS}}$ scheme at the renormalization scale, $\mu_{R} = 10^{15}$\,GeV.
We also show the neutrino parameters used in the present analysis.}
\label{tab:parameters}
\begin{center}
\renewcommand{\arraystretch}{1.2}
\begin{tabular}{|c|c|c|c|c|}  \hline
    $y_{u}/10^{-6}$ & $y_{c}/10^{-3}$ & $y_{t}$ & $y_{d}/10^{-6}$ & $y_{s}/10^{-4}$ \\ \hline 
    $2.83 \pm 0.35$ & $1.43 \pm 0.04$ & $0.436 \pm 0.003$ & $6.32 \pm 0.41$ & $1.26 \pm 0.07$ \\ \hline \hline
    $y_{b}/10^{-3}$ & $\sin{\theta_{12}^{\CKM}}$ & $\sin{\theta_{23}^{\CKM}}/10^{-2}$ & $\sin{\theta_{13}^{\CKM}}/10^{-3}$ & $\delta_{\mathrm{13}}^{\CKM}$ \\ \hline
    $6.01 \pm 0.06$ & $0.225 \pm 0.001$ & $4.67 \pm 0.09$ & $4.12 \pm 0.12$ & $1.14 \pm 0.03$ \\ \hline \hline
    $y_{e}/10^{-6}$ & $y_{\mu}/10^{-4}$ & $y_{\tau}/10^{-3}$ & $(\Delta m_{12}^{2}/\Delta m_{13}^{2})/10^{-2}$ & $\Delta m_{31}^{2}/10^{-3}\, \mathrm{eV}^{2}$ \\ \hline
    $2.718 \pm 0.004$ & $5.74 \pm 0.01$ & $9.70 \pm 0.02$ & $2.96 \pm 0.09$ & $2.507 \pm 0.027$ \\ \hline \hline
    $\sin^{2}{\theta_{12}^{\PMNS}}$ & $\sin^{2}{\theta_{23}^{\PMNS}}$ & $\sin^{2}{\theta_{13}^{\PMNS}}/10^{-2}$ & $\delta_{\mathrm{13}}^{\PMNS}$ & \\ \hline
    $0.304 \pm 0.012$ & $0.450 \pm 0.018$ & $2.25 \pm 0.01$ & $4.01 \pm 0.52$ & \\ \hline
  \end{tabular}
 \end{center}
\end{table}

\section{Matrix Decomposition and Integral Measure}
\label{sec:measure}

According to \cite{Haba:2000be}, 
we summarize the matrix decomposition and integral measure which we have used in this work.
The singular value decomposition of a complex 3$\,\times\,$3 matrix $A$ 
into a positive semi-definite diagonal matrix $\Sigma$ is given by,
\begin{align}
\label{eq:singular}
    A = U_{L} \Sigma U_{R}^{\dagger}\ , \quad \Sigma = \mathrm{diag}(\sigma_{1}, \sigma_{2}, \sigma_{3})\ ,
\end{align}
with some 3$\,\times\,$3 unitary matrices $U_L$ and $U_R$.
Then, the volume element of $A$
can be rewritten by
\begin{align}
\label{eq:element}
\prod_{i,j} d\, (\Re \, A_{ij})\, d\, (\Im \, A_{ij}) =
(\sigma_{1}^{2} - \sigma_{2}^{2})^{2} \, (\sigma_{2}^{2} - \sigma_{3}^{2})^{2} \, (\sigma_{3}^{2} - \sigma_{1}^{2})^{2} \,
d\sigma_{1}^{2} \, d\sigma_{2}^{2} \, d\sigma_{3}^{2} 
\frac{dU_{L} \, dU_{R}}{d\phi_{1} \, d\phi_{2} \, d\phi_{3}}\ ,
\end{align}
where $dU$'s are the invariant Haar measure over the U(3) group.
Here, 
we take
\begin{align}
\label{eq:Umatrix}
  U= e^{i\eta} e^{i\rho_{1} \lambda_{3} + i \rho_{2} \lambda_{8}}
   \begin{pmatrix}
  	1 & 0 & 0\\
	0 & c_{23} & s_{23}\\
	0 & -s_{23} & c_{23}
  \end{pmatrix} 
  \begin{pmatrix}
  	c_{13} & 0 & s_{13} e^{-i\delta_{\mathrm{13}}}\\
	0 & 1 & 0\\
	-s_{13}e^{i\delta_{\mathrm{13}}} & 0 & c_{13}
  \end{pmatrix}
  \begin{pmatrix}
  	c_{12} & s_{12} & 0\\
	-s_{12} & c_{12} & 0\\
	0 & 0 & 1 \end{pmatrix}
  e^{i\chi_{1} \lambda_{3} + i \chi_{2} \lambda_{8}},
\end{align}
with $\lambda_{3} = {\rm diag}(1, -1, 0)$ and $\lambda_{8} = {\rm 
diag} (1, 1, -2)/\sqrt{3}$ being the Gell-Mann matrices.  
Here, $c_{ij}=\cos\theta_{ij}$, $s_{ij}=\sin\theta_{ij}$ and $\theta_{12}$, $\theta_{23}$, and $\theta_{13}$ are Euler angles 
in $[0,\pi/2]$.
The CP-phases are given by 
$\eta$, $\rho_{1,2}$, $\chi_{1,2}$ 
and $\delta_{\mathrm{13}}$ are in $[0,2\pi)$.

For this parameterization, 
the Haar measure is given by
\begin{equation}
	d U = d s_{12}^{2} \, d c_{13}^{4} \, d s_{23}^{2} \, d\delta_{\mathrm{13}} \,
	d\eta \, d\rho_{1} \, d\rho_{2} \, d\chi_{1} \, d\chi_{2}\ ,
\end{equation}
where we omit the 
universal normalization since 
it is irrelevant for the Bayes inference.
The division by $(d\phi_1 \, d\phi_2 \, d\phi_3)$ in Eq.\,\eqref{eq:element} 
eliminates the redundancy 
of the common phase factor
associated with 
a phase rotation
 $U_L \rightarrow U_L \Phi$ and $U_R \rightarrow U_R \Phi$ where $\Phi=\mathrm{diag}(e^{i \phi_1},e^{i \phi_2},e^{i \phi_3} )$ for the relation in Eq.\,\eqref{eq:singular}.
 Concretely, we fix $\eta$, $\rho_{1,2}$ of $U_R$ to be $0$ in our numerical analysis.

The singular value decomposition of a complex 3$\,\times\,$3 symmetric matrix $B$ into 
 a positive semi-definite diagonal matrix $\Sigma$ is given by,
\begin{align}
B = U \Sigma U^{T}\ , \quad 
\Sigma =\mathrm{diag}(\sigma_1, \sigma_2, \sigma_3)\ , 
\end{align}
with a 3$\,\times\,$3 unitary matrix $U$.
Then the volume element of $B$
can be rewritten by
\begin{align}
\prod_{i,j} d\, (\Re \, B_{ij})\, d\, (\Im \, B_{ij})
= 
(\sigma_{1}^{2} - \sigma_{2}^{2}) \, (\sigma_{2}^{2} - \sigma_{3}^{2}) \, (\sigma_{3}^{2} - \sigma_{1}^{2}) \, d\sigma_{1}^{2} \, d\sigma_{2}^{2} \, d\sigma_{3}^{2} \, dU \ .
\end{align}
where $dU$ is the invariant Haar measure over the U(3) group.

\section{Details of Priors and Likelihoods}
\label{sec:measure_yukawa}

We write down the prior distributions, integral measures and likelihood functions used in our calculation.
The prior distribution functions for
$\kappa_{u,d,e,D}$ and $\kappa_{R,W}$ are the following forms respectively,
\begin{align}
\label{eq:pi_udeD}
\pi(\kappa^{(F)}) 
&= 
\pi(y^{(F)} \circ \epsilon^{-q^{(F)}}) 
= 
\left(\frac{1}{\sqrt{2\pi}\sigma} \right)^{18}
\exp\left({{ - \frac{\tr{[(y^{(F)} \circ \epsilon^{-q^{(F)}})^{\dagger} 
(y^{(F)} \circ \epsilon^{-q^{(F)}})]}}{2\sigma^2}}}\right)\ , \quad \nonumber \\
y^{(F)} 
&= 
U_{FL}\, \hat{y}^{(F)}\, U_{FR}^{\dagger}\ ,
\end{align}
for $F=\{u,d,e,D\}$, and 
\begin{align}
\label{eq:pi_RW}
\pi(\kappa^{(F)}) 
&= 
\pi(y^{(F)} \circ \epsilon^{-q^{(F)}}) 
= 
\left(\frac{1}{\sqrt{2\pi}\sigma} \right)^{6}
\left(\frac{1}{\sqrt{\pi}\sigma} \right)^{6}
\exp\left({{ - \frac{\tr{[(y^{(F)} \circ \epsilon^{-q^{(F)}})^{\dagger} 
(y^{(F)} \circ \epsilon^{-q^{(F)}})]}}{2\sigma^2}}}\right)\ , \quad \nonumber \\
y^{(F)} 
&= 
U_{F}\, \hat{y}^{(F)}\, U_{F}^{T}\ .
\end{align}
for $F=\{R,W\}$.
Here, 
the matrices with $\hat{\phantom{y}}$ are the positive semi-definite diagonal matrices.
We have used the Hadamard products in Eqs.\,\eqref{eq:hadamard}, \eqref{eq:hadamard_nu} and \eqref{eq:kappaW}.

The integral measures 
are the following forms respectively:
\begin{gather}
d\kappa^{(F)}
= 
J_F \, dy^{(F)}
= 
J_F\,
f(\hat{y}_{1}^{(F)}, \hat{y}_{2}^{(F)}, \hat{y}_{3}^{(F)})\,
d\, (\hat{y}_{1}^{(F)})^2\, d\,(\hat{y}_{2}^{(F)})^2 \, d\,(\hat{y}_{3}^{(F)})^2\,
\frac{dU_{FL}\, dU_{FR}}{d\phi_{1}\, d\phi_{2}\, d\phi_{3}}\ , \nonumber \\
f(\hat{y}_{1}^{(F)}, \hat{y}_{2}^{(F)}, \hat{y}_{3}^{(F)})
= 
\left[(\hat{y}_{1}^{(F)})^2 - (\hat{y}_{2}^{(F)})^2 \right]^2 \,
\left[(\hat{y}_{2}^{(F)})^2 - (\hat{y}_{3}^{(F)})^2 \right]^2 \,
\left[(\hat{y}_{3}^{(F)})^2 - (\hat{y}_{1}^{(F)})^2 \right]^2 \, \nonumber \\
\label{eq:measure_udeD}
J_F
= 
\prod_{1 \le i,j \le 3} \epsilon^{-2q^{(F)}_{ij}}\ ,
\end{gather}
for $F=\{u,d,e,D\}$, and 
\begin{gather}
\label{eq:measure_RW}
d\kappa^{(F)}
= 
J_F \, dy^{(F)}
= 
J_F\,
f(\hat{y}_{1}^{(F)}, \hat{y}_{2}^{(F)}, \hat{y}_{3}^{(F)})\,
d\,(\hat{y}_{1}^{(F)})^2\, d\,(\hat{y}_{2}^{(F)})^2 \, d\,(\hat{y}_{3}^{(F)})^2\,
dU_F\ , \nonumber \\
f(\hat{y}_{1}^{(F)}, \hat{y}_{2}^{(F)}, \hat{y}_{3}^{(F)})
= 
\left[(\hat{y}_{1}^{(F)})^2 - (\hat{y}_{2}^{(F)})^2 \right] \,
\left[(\hat{y}_{2}^{(F)})^2 - (\hat{y}_{3}^{(F)})^2 \right] \,
\left[(\hat{y}_{3}^{(F)})^2 - (\hat{y}_{1}^{(F)})^2 \right] \, \nonumber \\
J_F
= 
\prod_{1 \le i \le j \le 3} \epsilon^{-2q^{(F)}_{ij}}\ ,
\end{gather}
for $F=\{R,W\}$.
However, in the calculation of the seesaw mechanism case, 
we have performed a variable transformation from $y^{(R)}$ to $y^{(S)}$ obtained from Eq.\eqref{eq:seesaw_dim5} 
and integrated with respect to $y^{(S)}$.	
For this variable transformation, 
we refer to \cite{Fortin:2016zyf}, 
and the result is as follows:
\begin{align}
d \kappa^{(D)} \, d \kappa^{(R)}
= J_D\, J_R\, d y^{(D)}\, d y^{(R)} 
= J_D\, J_R\, 
\left|\frac{\det y^{(D)}}{\det y^{(S)}} \right|^{\beta(N+1)}\, 
d y^{(D)} d y^{(S)}\, ,
\end{align}
where $N$ represents the number of flavors, and in this work, $N=3$.
$\beta = 1\, (2)$ corresponds to real (complex) matrix elements, and in this work, $\beta=2$.

By using the above prior distributions and the likelihood functions,
the marginalized likelihood of a model $M$ is given by,
\begin{align}
\calZ
= \mathrm{P}(\mathbf{D}|M)
&= 
\int \mathrm{P} (\mathbf{D}|\mathbf{\Theta}, M) \mathrm{P} (\mathbf{\Theta}|M)\, d\mathbf{\Theta}  
= 
\int \mathcal{L}(\mathbf{x}) \,
\Bigg(\prod_{F}\, \pi(y^{(F)} \circ \epsilon^{-q^{(F)}}) \, 
J_F \, dy^{(F)}\Bigg) \,
d\epsilon \ .
\end{align}
Here, in the case of the dimension-five operator, $F=\{u,d,e,W\}$,
and in the case of the seesaw mechanism, $F=\{u,d,e,D,R\}$.
Due to the $\delta$--function likelihood,
the integration over $\hat{y}^{(F = u,d,e)}$ are trivially done, 
which replaces the values of $\hat{y}^{(F = u,d,e)}$ 
in Eqs.\,\eqref{eq:pi_udeD} and \eqref{eq:measure_udeD} with the observed values in Tab.\,\ref{tab:parameters}.
The integration with respect to $U_{uL}$ ($U_{eL}$) is converted to an integration over the CKM (PMNS) parameters by the relation $U_{uL} = U_{dL} V_{\CKM}^T$ ($U_{eL} = U_{\nu} V_\PMNS^T$).

For quark case:
\begin{align}
\calZ_{\mathrm{quark}}
&= 
\int 
\calL(\hat{y}^{(u)}_1) \calL(\hat{y}^{(u)}_2) \calL(\hat{y}^{(u)}_3)
\calL(\hat{y}^{(d)}_1) \calL(\hat{y}^{(d)}_2) \calL(\hat{y}^{(d)}_3)
\calL(s^{\CKM}_{12}) \calL(s^{\CKM}_{23}) \calL(c^{\CKM}_{13})
\calL(\delta^{\CKM}_{\mathrm{CP}}) \nonumber \\
&\times
\pi(y^{(u)} \circ \epsilon^{-q^{(u)}})\,
\pi(y^{(d)} \circ \epsilon^{-q^{(d)}})\,
J_u\,
f(\hat{y}_{1}^{(u)}, \hat{y}_{2}^{(u)}, \hat{y}_{3}^{(u)})\,
J_d\,
f(\hat{y}_{1}^{(d)}, \hat{y}_{2}^{(d)}, \hat{y}_{3}^{(d)})\, \nonumber \\
&\times
d\,(\hat{y}_{1}^{(u)})^2\, d\,(\hat{y}_{2}^{(u)})^2 \, 
d\,(\hat{y}_{3}^{(u)})^2\,
d\,(\hat{y}_{1}^{(d)})^2\, d\,(\hat{y}_{2}^{(d)})^2 \, 
d\,(\hat{y}_{3}^{(d)})^2\, 
\frac{dU_{uL}\, dU_{uR}}{d\phi_{1u}\, d\phi_{2u}\, d\phi_{3u}}\,
\frac{dU_{dL}\, dU_{dR}}{d\phi_{1d}\, d\phi_{2d}\, d\phi_{3d}}\,
d\epsilon \ .
\end{align}
The integration with respect to $U_{uL}$ is converted to an integration over the CKM parameters by the relation $U_{uL} = U_{dL} V_{\CKM}^T$.
Therefore, 
\begin{align}
\calZ_{\mathrm{quark}}
&= 
\int 
\calL(\hat{y}^{(u)}_1) \calL(\hat{y}^{(u)}_2) \calL(\hat{y}^{(u)}_3)
\calL(\hat{y}^{(d)}_1) \calL(\hat{y}^{(d)}_2) \calL(\hat{y}^{(d)}_3)
\calL(s^{\CKM}_{12}) \calL(s^{\CKM}_{23}) \calL(c^{\CKM}_{13})
\calL(\delta^{\CKM}_{\mathrm{CP}}) \nonumber \\
&\times
\pi(y^{(u)} \circ \epsilon^{-q^{(u)}})\,
\pi(y^{(d)} \circ \epsilon^{-q^{(d)}})\,
J_u\,
f(\hat{y}_{1}^{(u)}, \hat{y}_{2}^{(u)}, \hat{y}_{3}^{(u)})\,
J_d\,
f(\hat{y}_{1}^{(d)}, \hat{y}_{2}^{(d)}, \hat{y}_{3}^{(d)})\, \nonumber \\
&\times
d\,(\hat{y}_{1}^{(u)})^2\, d\,(\hat{y}_{2}^{(u)})^2 \, 
d\,(\hat{y}_{3}^{(u)})^2\,
d\,(\hat{y}_{1}^{(d)})^2\, d\,(\hat{y}_{2}^{(d)})^2 \, 
d\,(\hat{y}_{3}^{(d)})^2\, 
\frac{dV_{\CKM}\, dU_{uR}}{d\phi_{1u}\, d\phi_{2u}\, d\phi_{3u}}\,
\frac{dU_{dL}\, dU_{dR}}{d\phi_{1d}\, d\phi_{2d}\, d\phi_{3d}}\,
d\epsilon \ .
\end{align}
For dimension-five operator case:
\begin{align}
\calZ_{\dimfive}
&= 
\int 
\calL(\hat{y}^{(e)}_1) \calL(\hat{y}^{(e)}_2) \calL(\hat{y}^{(e)}_3)
\calL(r)
\calL(s^{\PMNS}_{12}) \calL(s^{\PMNS}_{23}) \calL(c^{\PMNS}_{13})
\calL(\delta^{\PMNS}_{\mathrm{CP}}) \nonumber \\
&\times
\pi(y^{(e)} \circ \epsilon^{-q^{(e)}})\,
\pi(y^{(W)} \circ \epsilon^{-q^{(W)}})\,
J_e\,
f(\hat{y}_{1}^{(e)}, \hat{y}_{2}^{(e)}, \hat{y}_{3}^{(e)})\,
J_W\,
f(\hat{y}_{1}^{(W)}, \hat{y}_{2}^{(W)}, \hat{y}_{3}^{(W)})\, \nonumber \\
&\times
d\,(\hat{y}_{1}^{(e)})^2\, d\,(\hat{y}_{2}^{(e)})^2 \, 
d\,(\hat{y}_{3}^{(e)})^2\,
d\,(\hat{y}_{1}^{(W)})^2\, d\,(\hat{y}_{2}^{(W)})^2 \, 
d\,(\hat{y}_{3}^{(W)})^2\, 
\frac{dU_{eL}\, dU_{eR}}{d\phi_{1e}\, d\phi_{2e}\, d\phi_{3e}}\,
dU_W\,
d\epsilon \ .
\end{align}
The integration with respect to $U_{eL}$ is converted to an integration over the CKM parameters by the relation $U_{eL} = U_{F} V_{\PMNS}^T$.
The integration over $\hat{y}^{(W)}$'s 
is rearranged by 
\begin{align}
d (\hat{y}^{(W)}_{1})^{2} \, 
d(\hat{y}^{(W)}_{2})^{2} \, d(\hat{y}^{(W)}_{3})^{2}
= 
d(\hat{y}^{(W)}_{1})^{2} \, d (\Delta \hat{y}^{2}_{21}) \, 
d(\Delta \hat{y}^{2}_{31})
= 
\Delta\hat{y}_{31}^{2}\,
d(\hat{y}^{(W)}_{1})^{2} \, 
d(\Delta\hat{y}_{31}^{2})\,
dr \ ,
\end{align}
where $r = \Delta y_{21}^2/\Delta y_{31}^2$.
Note that $\Delta\hat{y}_{31}^{2}\ge 0$ 
and $r>0$.
Additionally, 
\begin{align}
f(\hat{y}_{1}^{(F)}, \hat{y}_{2}^{(F)}, \hat{y}_{3}^{(F)})
&= 
\left[(\hat{y}_{1}^{(F)})^2 - (\hat{y}_{2}^{(F)})^2 \right] \,
\left[(\hat{y}_{2}^{(F)})^2 - (\hat{y}_{3}^{(F)})^2 \right] \,
\left[(\hat{y}_{3}^{(F)})^2 - (\hat{y}_{1}^{(F)})^2 \right] \nonumber \\
&= 
-\Delta y^{2}_{21}\, (\Delta y^{2}_{21} - \Delta y^{2}_{31})\, 
\Delta y^{2}_{31}
= (\Delta y^{2}_{31})^3\, r\, (1-r)
\end{align}
Therefore, 
\begin{align}
\calZ_{\dimfive}
&= 
\int 
\calL(\hat{y}^{(e)}_1) \calL(\hat{y}^{(e)}_2) \calL(\hat{y}^{(e)}_3)
\calL(r)
\calL(s^{\PMNS}_{12}) \calL(s^{\PMNS}_{23}) \calL(c^{\PMNS}_{13})
\calL(\delta^{\PMNS}_{\mathrm{CP}}) \nonumber \\
&\times
\pi(y^{(e)} \circ \epsilon^{-q^{(e)}})\,
\pi(y^{(W)} \circ \epsilon^{-q^{(W)}})\,
J_e\,
f(\hat{y}_{1}^{(e)}, \hat{y}_{2}^{(e)}, \hat{y}_{3}^{(e)})\,
J_W\,
(\Delta y^{2}_{31})^4\, r\, (1-r)\, \nonumber \\
&\times
d\,(\hat{y}_{1}^{(e)})^2\, d\,(\hat{y}_{2}^{(e)})^2 \, 
d\,(\hat{y}_{3}^{(e)})^2\,
d(\hat{y}^{(W)}_{1})^{2} \, 
d(\Delta\hat{y}_{31}^{2})\,
dr\,
\frac{dV_{\PMNS}\, dU_{eR}}{d\phi_{1e}\, d\phi_{2e}\, d\phi_{3e}}\,
dU_W\,
d\epsilon \ .
\end{align}
For seesaw mechanism case:
\begin{align}
\calZ_{\seesaw}
&= 
\int 
\calL(\hat{y}^{(e)}_1) \calL(\hat{y}^{(e)}_2) \calL(\hat{y}^{(e)}_3)
\calL(r)
\calL(s^{\PMNS}_{12}) \calL(s^{\PMNS}_{23}) \calL(c^{\PMNS}_{13})
\calL(\delta^{\PMNS}_{\mathrm{CP}}) \nonumber \\
&\times
\pi(y^{(e)} \circ \epsilon^{-q^{(e)}})\,
\pi(y^{(D)} \circ \epsilon^{-q^{(D)}})\,
\pi(y^{(R)} \circ \epsilon^{-q^{(R)}})\, \nonumber \\
&\times
J_e\, J_{D}\, J_R\,
d y^{(e)} d y^{(D)} d y^{(R)}
d\epsilon \ .
\end{align}
By transforming $y^{(R)}$ into $y^{(S)}$, 
\begin{align}
\calZ_{\seesaw}
&= 
\int 
\calL(\hat{y}^{(e)}_1) \calL(\hat{y}^{(e)}_2) \calL(\hat{y}^{(e)}_3)
\calL(r)
\calL(s^{\PMNS}_{12}) \calL(s^{\PMNS}_{23}) \calL(c^{\PMNS}_{13})
\calL(\delta^{\PMNS}_{\mathrm{CP}}) \nonumber \\
&\times
\pi(y^{(e)} \circ \epsilon^{-q^{(e)}})\,
\pi(y^{(D)} \circ \epsilon^{-q^{(D)}})\,
\pi(y^{(S)}, y^{(D)}, \epsilon^{-q^{(R)}} )\, \nonumber \\
&\times
J_e\, J_{D}\, J_R\,
\left|\frac{\det y^{(D)}}{\det y^{(S)}} \right|^{8}\,
d y^{(e)}\, d y^{(D)}\, d y^{(S)}\,
d\epsilon \ .
\end{align}
Here, $\pi(y^{(S)}, y^{(D)}, \epsilon^{-q^{(R)}} )$ is a function of $y^{(S)}$, $y^{(D)}$ and $\epsilon^{-q^{(R)}}$.
Therefore, 
\begin{align}
\calZ_{\seesaw}
&= 
\int 
\calL(\hat{y}^{(e)}_1) \calL(\hat{y}^{(e)}_2) \calL(\hat{y}^{(e)}_3)
\calL(r)
\calL(s^{\PMNS}_{12}) \calL(s^{\PMNS}_{23}) \calL(c^{\PMNS}_{13})
\calL(\delta^{\PMNS}_{\mathrm{CP}}) \nonumber \\
&\times
\pi(y^{(e)} \circ \epsilon^{-q^{(e)}})\,
\pi(y^{(D)} \circ \epsilon^{-q^{(D)}})\,
\pi(y^{(S)}, y^{(D)}, \epsilon^{-q^{(R)}} )\, \nonumber \\
&\times
J_e\,
f(\hat{y}_{1}^{(e)}, \hat{y}_{2}^{(e)}, \hat{y}_{3}^{(e)})\,
J_D\,
f(\hat{y}_{1}^{(D)}, \hat{y}_{2}^{(D)}, \hat{y}_{3}^{(D)})\,
J_R\,
(\Delta y^{2}_{31})^4\, r\, (1-r)\, 
\left|\frac{\det y^{(D)}}{\det y^{(S)}} \right|^{8}\, \nonumber \\
&\times
d\,(\hat{y}_{1}^{(e)})^2\, d\,(\hat{y}_{2}^{(e)})^2 \, 
d\,(\hat{y}_{3}^{(e)})^2\,
d\,(\hat{y}_{1}^{(D)})^2\, d\,(\hat{y}_{2}^{(D)})^2 \, 
d\,(\hat{y}_{3}^{(D)})^2\,
d(\hat{y}^{(S)}_{1})^{2} \, 
d(\Delta\hat{y}_{31}^{2})\,
dr\, \nonumber \\
&\times
\frac{dV_{\PMNS}\, dU_{eR}}{d\phi_{1e}\, d\phi_{2e}\, d\phi_{3e}}\,
\frac{dU_{DL}\, dU_{DR}}{d\phi_{1D}\, d\phi_{2D}\, d\phi_{3D}}\,
dU_S\,
d\epsilon \ .
\end{align}

\clearpage
\section{Bayes Factor in Inverted Ordering}
\label{sec:inverted_comment}

\begin{figure}[h!]
    \centering
    \begin{subfigure}[b]{0.49\textwidth}
        \centering        \includegraphics[width=\textwidth]{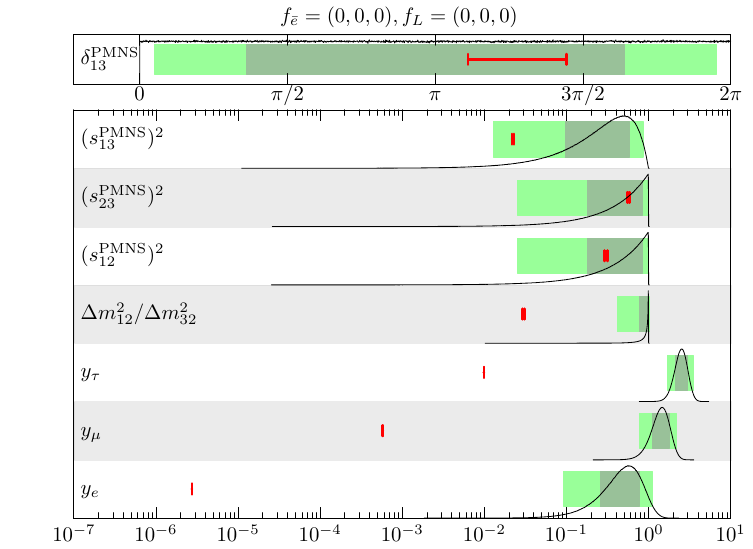}
        \caption{Seesaw mechanism.}
        \label{fig:seesawIO_0}
    \end{subfigure}
    \hfill
    \begin{subfigure}[b]{0.49\textwidth}
        \centering
        \includegraphics[width=\textwidth]{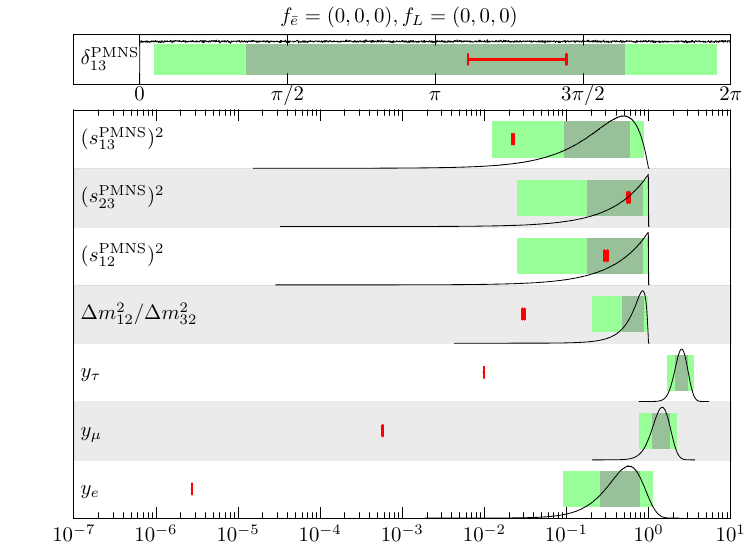}
        \caption{Dimension-five operator.}
        \label{fig:dim5IO_0}
    \end{subfigure}
    \caption{Figures (a) and (b) present the predictions for the inverted ordering from $\order{1}$ distributions of $\kappa$'s (without FN charges) for the seesaw mechanism and the dimension-five operator, respectively.
The meanings of the red bars and band colors are the same as in Fig.\,\ref{fig:conv}.
    }
    \label{fig:IO_0}
\end{figure}

We have primarily discussed the normal ordering, which is favored over the inverted ordering ($m_3 < m_1 < m_2$) based on experimental results with SK data\,\cite{Esteban:2020cvm}. 
In this appendix, 
we briefly address the case of the inverted ordering.
First, we present the predictions of the PMNS parameters for the inverted ordering without FN charges in the case of the seesaw mechanism and the dimension-five operator in Fig.\,\ref{fig:seesawIO_0} and Fig.\,\ref{fig:dim5IO_0}, respectively. 
These figures demonstrate that the ratio of mass differences, 
${\Delta}m_{12}^2 / \Delta m_{32}^2$, 
significantly deviates from the observed value.
This result should be compared with the results without FN charges for the normal ordering shown in Fig.\,\ref{fig:conv}. 
Concretely, the Bayes factor for the neutrino sector for the seesaw mechanism,
\begin{align}
\calZ_{0,\seesaw,
\nu}^{\mathrm{IO}}
&= \calZ^{\mathrm{IO}}_{0,\seesaw,\nu\mathchar`-\mathrm{mass}} \times \calZ_{0,\mathrm{PMNS}} 
\simeq 1.7\times 10^{-3}\ ,
\end{align} 
where
$\calZ_{0,\mathrm{PMNS}}\simeq 0.31$
is given in Eq.\,\eqref{eq:ZPMNS}
and $\calZ^{\mathrm{IO}}_{0,\seesaw,\nu\mathchar`-\mathrm{mass}}\simeq 0.0056$.
For the dimension-five operator case is given by,
\begin{align}
\calZ_{0,\dimfive,\nu}^{\mathrm{IO}}
&= \calZ^{\mathrm{IO}}_{0,\dimfive,\mathrm{neutrino}} \times \calZ_{0,\mathrm{PMNS}} 
\simeq 9.6\times 10^{-3}\ .
\end{align} 
where we have utilized,
\begin{align}
\calZ^{\mathrm{IO}}_{0,\dimfive,\mathrm{neutrino}}
&= \frac{512 (-1+r) r}{(-2+r)^5}
= 0.031\ , 
\end{align}
with $r= \Delta m^2_{12}/\Delta m^2_{32}$.

This deviation in the mass difference ratio for the inverted ordering cannot be resolved in the seesaw mechanism by assigning FN charges to the charged lepton sector as long as the FN charges of the right-handed neutrinos are fixed at $f_{\Nbar,i} = 0$. 
In such a case, 
the Bayes factor of the neutrino sector for the inverted ordering does not become large.
On the other hand, 
it has been pointed out that if FN charges are assigned to the right-handed neutrinos, the observed values of the neutrino parameters can be well reproduced even in the case of an inverted ordering \cite{King:2000ce}. 
However, 
we do not pursue this possibility in this paper.

\begin{table}[h!]
 \begin{center}
   \caption{The six FN charge assignments of the lepton sector with larger Bayes factors for the inverted ordering in the case of the dimension-five operator. 
We normalize the marginal likelihood 
by that for the normal ordering without FN charges.
The range of $\epsilon$ is the 95\% CI of its posterior distribution.}
\begin{tabular}[t]{|c|c|c|c|} \hline
    $f_L$ & $f_{\ebar}$ & $\mathrm{log}_{10} (\calZ^{\mathrm{IO}}_{\dimfive}/\calZ_{0,\dimfive})$ & $\epsilon$ \\ \hline
$3,-6,3$ & $-10,1,-6$ & $53.57\pm0.02$ & $0.174 \to 0.212$ \\
$-8,-4,2$ & $-7,1,6$ & $53.14\pm0.03$ & $0.067 \to 0.091$ \\
$4,-2,4$ & $7,-1,-2$ & $53.12\pm0.02$ & $0.066 \to 0.092$ \\
$-5,2,8$ & $10,1,-10$ & $52.81\pm0.03$ & $0.077 \to 0.106$ \\
$-1,1,1$ & $-8,-5,-4$ & $52.78\pm0.02$ & $0.183 \to 0.216$ \\
$3,-8,4$ & $-10,2,-7$ & $52.59\pm0.02$ & $0.195 \to 0.242$\\
    \hline
  \end{tabular}
\label{tab:IO_dim5}
 \end{center} 
\end{table}
\begin{figure}[h!]
    \centering
    \begin{subfigure}[b]{0.49\textwidth}
        \centering
        \includegraphics[width=\textwidth]{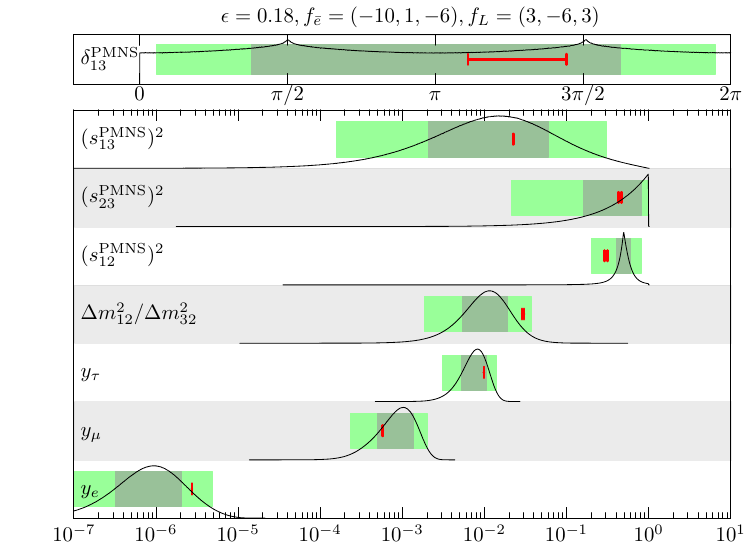}
        \caption{}
        \label{fig:dim5IO_charge1}
    \end{subfigure}
    \hfill
    \begin{subfigure}[b]{0.49\textwidth}
        \centering
        \includegraphics[width=\textwidth]{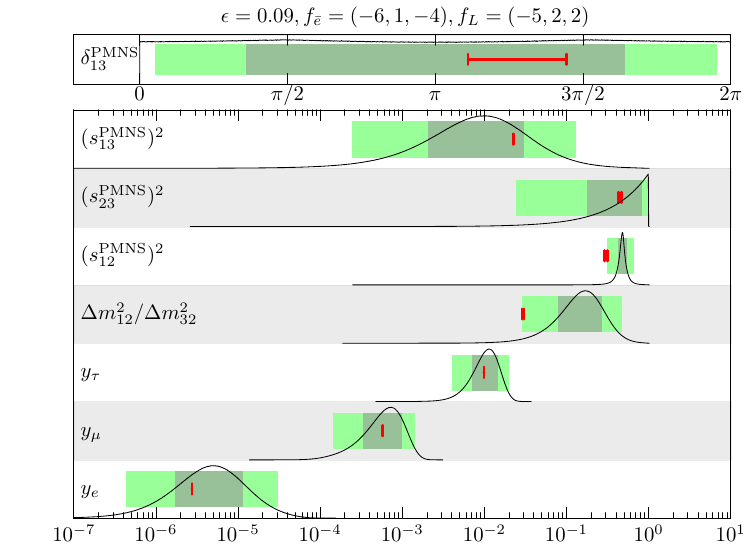}
        \caption{}
        \label{fig:dim5IO_charge2}
    \end{subfigure}
    \caption{Figures (a) and (b) present predictions from $\order{1}$ distributions of $\kappa$’s with FN charges, which exhibit large Bayes factors, for the inverted ordering in the dimension-five operator. 
   The meanings of the red bars and band colors are the same as in Fig.\,\ref{fig:conv}.
}
  \label{fig:IO_bests}
\end{figure}

\begin{figure}[t!]
	\centering
	{\includegraphics[width=0.7\textwidth]{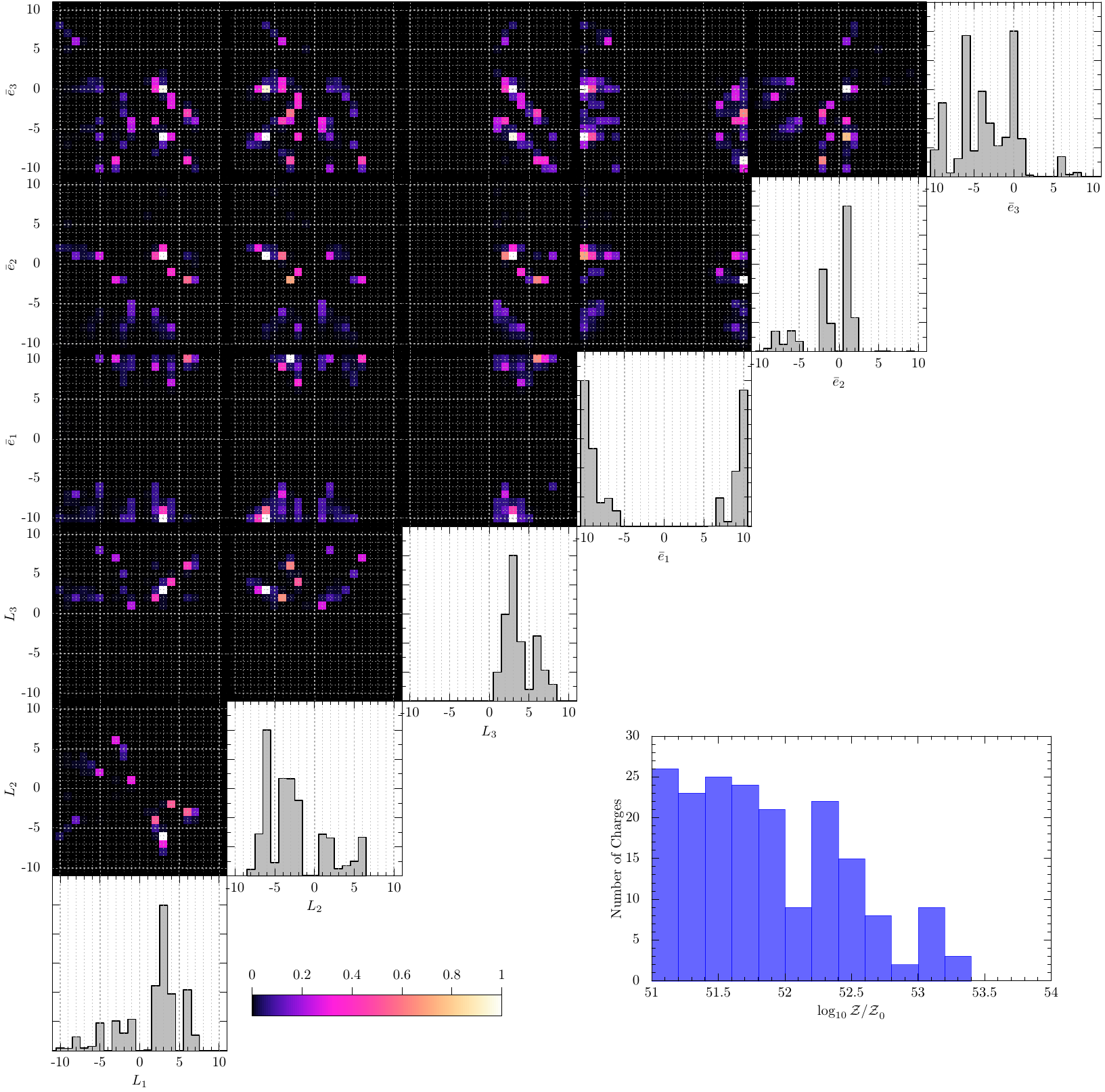}}
 \caption{The upper-left panel shows a two-dimensional histogram of our the lepton charges in the inverted ordering. 
 These results are normalized after being weighted by the values of the Bayes factor.
 The lower-right panel shows the distribution of Bayes factors of the lepton FN charge candidates.}
 \label{fig:io_corner}
\end{figure}

In contrast, 
in the dimension-five operator case with FN charges, 
there exist favorable FN charge assignments that can reproduce the neutrino parameters for the inverted ordering. 
In Tab.\,\ref{tab:IO_dim5}, 
we present examples of FN charge assignments for the inverted ordering in the dimension-five operator case. 
Here, 
we normalize the marginalized likelihood by that for the normal ordering in the dimension-five operator case. 
As in the case of Sec.\,\ref{sec:result_charge},
we show the predictions of the lepton sector parameters for the two of best FN charge assignments in Fig.\,\ref{fig:IO_bests}.
In addition,
we also show a two-dimensional histogram of the FN charges in Fig.\,\ref{fig:io_corner}.
A caveat is that, 
in the calculation of the Bayes factor shown in the table, 
we do not account for the minimum value of the likelihood function $\mathit{\Delta}\chi^2_\mathrm{min}$ reported in Ref.\,\cite{Esteban:2020cvm}. 
If this lower bound is considered, 
the Bayes factor would be further suppressed by approximately,
$e^{-\mathit{\Delta}\chi^2_\mathrm{min}/2} \simeq 0.04$.

\FloatBarrier
\section{Derivation of \texorpdfstring{Eq.\,\eqref{eq:combined_evidence}}{}}
\label{app:combined_evidence}

We summarize the derivation of Eq.\,\eqref{eq:combined_evidence}.
First, marginalized likelihoods of the quark and lepton sectors are given by
\begin{align}
\calZ_{\mathrm{quark}} 
= 
\int d \epsilon\, d \mathbf{\Theta}_q\, \P_q(\epsilon, \mathbf{\Theta}_q)
= 
\calZ_{\mathrm{quark}}  \, \int d \epsilon\, \rho_q(\epsilon)\, ,\quad
\calZ_l
= 
\int d \epsilon\, d \mathbf{\Theta}_l\, \P_l(\epsilon, \mathbf{\Theta}_l)
= 
\calZ_l \, \int d \epsilon\, \rho_l(\epsilon)\, ,
\end{align}
where $\P_q(\epsilon, \mathbf{\Theta}_q)$ and $\P_l(\epsilon, \mathbf{\Theta}_l)$ are the product of likelihoods and prior distributions in the quark sector and the lepton sector, respectively. 
$\mathbf{\Theta}_q$ and $\mathbf{\Theta}_l$ represent model parameters other than $\epsilon$.
$\rho_q(\epsilon)$ and $\rho_l(\epsilon)$ are the posterior distribution of $\epsilon$ in the quark sector and the lepton sector, respectively.
In this time, using the above expressions, 
$\calZ_{\mathrm{C}}$ can be derived as follows: 
\begin{align}
\calZ_{\mathrm{C}} 
&=
\int d \epsilon \, 
d \mathbf{\Theta}_q\, d \mathbf{\Theta}_l\, \P_q(\epsilon, \mathbf{\Theta}_q)\, 
\P_l(\epsilon, \mathbf{\Theta}_l)
= 
\int d \epsilon_q\, d \epsilon_l\, 
d \mathbf{\Theta}_q\, d \mathbf{\Theta}_l\,
\delta(\epsilon_q - \epsilon_l) 
\P_q(\epsilon_q, \mathbf{\Theta}_q)\, 
\P_l(\epsilon_l, \mathbf{\Theta}_l) \nonumber \\
&=
\calZ_{\mathrm{quark}}  \, \calZ_l \, 
\int d \epsilon_q \, d \epsilon_l \,
\delta(\epsilon_q - \epsilon_l)
\rho_q(\epsilon_q)\, \rho_l(\epsilon_l)
=
\calZ_{\mathrm{quark}}  \, \calZ_l \, 
\int d \epsilon \rho_q(\epsilon)\, \rho_l(\epsilon)\, .
\end{align}

\bibliographystyle{apsrev4-1}
\bibliography{bibtex}

\end{document}